\documentclass[onecolumn,preprintnumbers,amsmath,amssymb,notitlepage]{revtex4}
\usepackage{graphicx}
\usepackage{epsfig}
\usepackage{amsmath}
\usepackage{amsfonts}
\usepackage{amssymb}
\usepackage{textcomp}
\usepackage{feynmf}

\begin{document}
\title{Canonical Noncommutativity Algebra for the Tetrad Field in General Relativity}

\author{Martin Kober}
\email{kober@fias.uni-frankfurt.de}
\email{kober@th.physik.uni-frankfurt.de}

\affiliation{Frankfurt Institute for Advanced Studies (FIAS),
Johann Wolfgang Goethe-Universit\"at,
Ruth-Moufang-Strasse 1, 60438 Frankfurt am Main, Germany}
\date{\today}

\begin{abstract}
General relativity under the assumption of noncommuting components of the tetrad field is considered in
this paper. Since the algebraic properties of the tetrad field representing the gravitational field are
assumed to correspond to the noncommutativity algebra of the coordinates in the canonical case of
noncommutative geometry, this idea is closely related to noncommutative geometry as well as to canonical
quantization of gravity. According to this presupposition generalized field equations for general relativity
are derived which are obtained by replacing the usual tetrad field by the tetrad field operator
within the actions and then building expectation values of the corresponding field equations between coherent
states. These coherent states refer to creation and annihilation operators created from the components of
the tetrad field operator. In this sense the obtained theory could be regarded as a kind of semiclassical
approximation of a complete quantum description of gravity. The consideration presupposes a special choice
of the tensor determining the algebra providing a division of space-time into two two-dimensional planes.
\end{abstract}

\maketitle

\section{Introduction}

The unification of quantum theory and general relativity is perhaps the most important question
in contemporary fundamental physics. One very important approach concerning a quantum theoretical
description of general relativity is the canonical quantization of the gravitational field.
In particular, the special manifestation given by loop quantum gravity \cite{Rovelli:1989za},\cite{Rovelli:1994ge}
based on the new variables to describe the gravitational field introduced by Ashtekar
\cite{Ashtekar:1986yd},\cite{Ashtekar:1987gu} is considered as a very promising candidate
for a quantum description of the gravitational field.
Noncommutative geometry represents another very promising concept towards a quantum theory of gravity.
In \cite{Seiberg:1999vs} has been shown how to maintain gauge invariance in the context of
noncommutative geometry, if the usual approach to noncommutative geometry with the
star product is presupposed. Various quantum field theories have been formulated by using the star product
\cite{Grosse:1995ar,Grosse:1996mz,Chaichian:1999wy,Chaichian:2004za,Gomis:2000gy,Amorim:2001gs,Calmet:2001na,
Calmet:2003jv,Calmet:2004yj,Calmet:2006zy,Calmet:2006nw,Denk:2004um,Melic:2005fm,Melic:2005am,Arean:2005ar,
Spallucci:2006zj,Acatrinei:2002sb,Malik:2003qa,Kossow:2006ym,Rosenbaum:2006yu,Tureanu:2006pb,Huang:2006tz,
Joung:2007qv,Kersting:2009vw,Balachandran:2009as,Popovic:2010zz}.
Also general relativity and modified gravity theories on noncommutative space-time
\cite{Chamseddine:1991qh,Chamseddine:1992yx,Mohammedi:1992ms,Madore:1993br,Sitarz:1994bd,Landi:1994cc,Hawkins:1996uu,
Chamseddine:2000zu,Vacaru:2000yk,Okawa:2000sh,Nishino:2001gt,Langmann:2001yr,Cacciatori:2002gq,Avramidi:2003mk,
Chamseddine:2003we,Yang:2004vd,Muthukumar:2004wj,Bourouaine:2005tt,Aschieri:2005yw,Valtancoli:2005js,Calmet:2005qm,
Calmet:2006iz,Buric:2006di,AlvarezGaume:2006bn,Mukherjee:2006nd,Szabo:2006wx,Harikumar:2006xf,
Aschieri:2005zs,Aschieri:2006kc,Zupnik:2005ph,Yang:2006dk,Banerjee:2007th,MullerHoissen:2007xy,Steinacker:2007dq,
Klammer:2008df,Steinacker:2008ri,Aschieri:2009ky,Vassilevich:2009cb,Asakawa:2009yb,Cortese:2010ze,Miao:2010kr,Miao:2010yq,
Nicolini:2010nb}
as well as aspects of cosmology in the context of noncommutative geometry
\cite{Lizzi:1995kq,Lizzi:2002ib,Barbosa:2004kp,Klammer:2009ku}
have been considered. The relation of noncommutative geometry and quantum gravity has been treated in 
\cite{Heller:1999zz,Heller:2000rr,Heller:2003gz,Heller:2005qa,Moffat:2000fv,Moffat:2000gr,Vassilevich:2004ym,
Freidel:2005me,Martinetti:2005ny,Martinetti:2009tr,Aastrup:2006ib,Aastrup:2010ds}.

In this paper a kind of combination of quantum gravity and noncommutative geometry is suggested.
General relativity is explored under the assumption that the components of the
gravitational field, or the tetrad field to be more specific, fulfil commutation relations which
correspond to the algebra of the canonical case of noncommutative geometry which usually refers to
the space-time coordinates. The relation to canonical quantum gravity consists in the postulated
canonical commutation relations for the gravitational field in the sense of a field quantization and
the relation to noncommutative geometry consists in the special form of the commutation relations
corresponding to the canonical case of noncommutative geometry, since in the approach of this paper
the commutation relations of noncommutative geometry are transferred from the coordinates to the
components of the tetrad field.  
Quantum field theory on noncommutative space-time in the sense of noncommuting space-time coordinates
is  usually treated by using the star product approach which is based on Weyl quantization and maps products
of fields depending on noncommuting coordinates to products of fields depending on usual coordinates.
Of course, this approach cannot be used with respect to the noncommutativity of the gravitational
field considered in this paper, since here the components of a field itself do not commute with
each other. This is in contrast to the usual formulation of quantum field theories on noncommutative
spacetime, where the components of the coordinates fulfil nontrivial commutation relations.
However, there exists another important approach to treat noncommutative geometry which is called the
coherent state approach, because there are defined coherent states which refer to creation and annihilation
operators obtained from linear combinations of the noncommuting quantities, namely the components of the
position vector in case of usual noncommutative geometry. In this approach the generalized quantities
depending on the noncommuting coordinates are mapped to generalized quantities depending on usual coordinates
by building expectation values between such coherent states.
The coherent state approach has been developed in \cite{Smailagic:2003rp},\cite{Smailagic:2003yb},\cite{Smailagic:2004yy}
where has been calculated an extended expression for plane waves depending on noncommuting coordinates.
In \cite{Kober:2010um} the coherent state approach has been extended to the case of noncommuting coordinates and momenta
and an extended expression for plane waves has also been determined, and using this expression the corresponding generalization of quantum field theory has finally been considered and derived a propagator.
Also thermodynamics has been treated within the coherent state approach to noncommutative geometry \cite{Huang:2008yy}.
In contrast to the star product the treatment of noncommutative geometry within the coherent state approach can be
transferred to the case of noncommuting fields and especially to the case of noncommuting components of the tetrad
field as they are treated in this paper.

The structure of this paper is as follows: A short review of the coherent state approach to noncommutative geometry
with respect to the usual case of noncommuting coordinates is given at the beginning.
Then the noncommutativity algebra of the tetrad field is provided. To enable the application of the
coherent state approach to noncommuting components of the tetrad field it is inevitable to use a special form of the
tensor defining the noncommutativity which implies a foliation of space-time into a couple of two-dimensional
submanifolds. Based on this algebra linear combinations of components of the tetrad field which behave like creation
and annihilation operators are defined and with respect to them coherent states referring indirectly to the tetrad
field can be introduced .
According to the considerations mentioned above expressions depending on the noncommuting gravitational
field will be mapped to expressions depending on the usual commuting gravitational field by building
expectation values between coherent states referring to the gravitational field. As a first application
of the developed concept, besides the expectation value of the resulting metric operator and the
tetrad field operator which is equal to the usual tetrad field itself, the expectation value of the
volume element with respect to the coherent states is calculated.
Using these concepts the generalized field equation for a matter field coupled
to the gravitational field and the generalized Einstein field equation are subsequently calculated,
which are obtained by replacing the usual tetrad field by the tetrad field operator obeying the noncommutativity
algebra within the corresponding actions, varying the actions with respect to the tetrad field operator and then
building the expectation value between coherent states.
A calculation to the third order in an expansion of the tetrad field around the Kronecker symbol
corresponding to the Minkowski metric in both cases is performed, because this is the lowest order leading to a
deformation of the Einstein field equation of the gravitational field.
This generalized description of the dynamics of the gravitational field according to general relativity
with the noncommutative tetrad field can be considered as a special kind of a semiclassical description
of general relativity.

\section{Review of the Coherent State Approach to Noncommutative Geometry}

In the coherent state approach to noncommutative geometry a canonical algebra between
the coordinates is assumed which reads as follows:

\begin{equation}
[\hat x^\mu,\hat x^\nu]=i\Theta^{\mu\nu},\quad \mu,\nu=1,...,D,
\end{equation}
where $[\hat A,\hat B]=\hat A \hat B-\hat B \hat A$ and $\Theta^{\mu\nu}$ is of the following special shape:

\begin{equation}
\Theta^{\mu\nu}={\rm diag}\left(\Theta_1,\Theta_2,...,\Theta_{D/2}\right),
\end{equation}
where the $\Theta_i$ are defined as

\begin{equation}
\Theta_i=\left(\begin{matrix}0 & \theta_i \\ -\theta_i & 0\end{matrix}\right),\quad i=1,...,d=D/2.
\label{Theta}
\end{equation}
This means that space-time is divided into two-dimensional submanifolds by the noncommutativity tensor.
If the components of the space-time coordinate are denoted as follows:

\begin{equation}
\hat x^\mu=\left(\hat x^1, \hat x^2,...,\hat x^{2d-1},\hat x^{2d}\right),
\end{equation} 
then because of ($\ref{Theta}$) it holds for the components of $\hat x^\mu$

\begin{equation}
\left[\hat x^{2i-1},\hat x^{2i}\right]=i\theta_i,\quad i=1,...,d.
\label{nc_planes}
\end{equation}
Creation and annihilation operators can now be defined according to

\begin{equation}
\hat a_i=\frac{1}{\sqrt{2\theta_i}}\left(\hat x_{2i-1}+i\hat x_{2i}\right)\quad,\quad
\hat a_i^{\dagger}=\frac{1}{\sqrt{2\theta_i}}\left(\hat x_{2i-1}-i\hat x_{2i}\right),\quad i=1,...,d,
\end{equation}
which fulfil the algebra

\begin{equation}
\left[\hat a_i,\hat a_j^{\dagger}\right]=\delta_{ij},\quad i,j=1,...,d,
\label{commutator_creation_annihilation}
\end{equation}
where $\delta_{ij}$ denotes the Kronecker symbol. Accordingly $d$ pairs of creation and annihilation operators are
defined which are constructed by the $d=D/2$ pairs of coordinates referring to the $d$ two-dimensional submanifolds
defined by ($\ref{nc_planes}$). With respect to ($\ref{commutator_creation_annihilation}$) coherent states can
be defined according to

\begin{equation}
|a \rangle=\prod_i \exp\left(-\frac{|a_i|^2}{2}\right)\exp\left(a_i \hat a_i^{\dagger}\right)|0\rangle
=\prod_i \exp\left(-\frac{|a_i|^2}{2}\right)\sum_{n=0}^{\infty}\frac{a_i^n}{\sqrt{n !}}
\frac{\hat a_i^{\dagger n}}{\sqrt{n !}}|0\rangle
=\prod_i \exp\left(-\frac{|a_i|^2}{2}\right)\sum_{n=0}^{\infty}\frac{a_i^n}{\sqrt{n !}}|n_{a_i}\rangle,
\label{coherent_states_ncg}
\end{equation}
where $a_i$ and $a_i^{*}$ describe the corresponding eigenvalues to the operators $\hat a_i$ and
$\hat a_i^{\dagger}$, the $|n_{ai}\rangle$ describe the eigenstates of the occupation number
operator $a_i^{\dagger} a_i$ and $|0\rangle$ denotes the vacuum state. For the coherent states
($\ref{coherent_states_ncg}$) hold eigenvalue equations which read

\begin{equation}
\hat a_i|a \rangle=a_i|a \rangle\quad,\quad \langle a|\hat a_i^{\dagger}=\langle a|a_i^{*}.
\end{equation}
By using these coherent states an expectation value of any function $f(\hat x)$ depending
on the noncommuting coordinates can be defined according to

\begin{equation}
\langle f(\hat x)\rangle=\langle a|f(\hat x_1,\hat x_2,...,\hat x_{2d-1},\hat x_{2d})|a\rangle
=\langle a|f(\hat a_1,\hat a_1^{\dagger},...,\hat a_{d},\hat a_d^{\dagger})|a\rangle=F(a_1,a_1^{*}...,a_{d},a_d^{*})=F(x),
\label{expectation_value_ncg}
\end{equation}
where $F(x)$ is a new function depending on the usual coordinates $x_i$ given by the expectation value of the
corresponding operators: $x_i=\langle a|\hat x_{i}|a\rangle$. This means that it is possible to
map arbitrary functions depending on noncommutative coordinates to functions (with additional terms)
depending on usual coordinates by using the coherent states and in this sense the coherent states
provide an alternative procedure besides the star product to map products of functions on
noncommutative space-time to expressions with additional terms on usual space-time. The coherent state
approach has been developed in \cite{Smailagic:2003rp},\cite{Smailagic:2003yb},\cite{Smailagic:2004yy}
and extended in \cite{Kober:2010um},\cite{Huang:2008yy}. If one refers to the usual case of a
(3+1)-dimensional space-time, one has to choose $d=2$ which means that space-time is divided into
two two-dimensional submanifolds. In the next section an analogue noncommutativity algebra for the tetrad
field will be introduced and after this the coherent state approach presented in this section will be
transferred to this case of noncommuting tetrad fields.

\section{Noncommutativity Algebra for the Tetrad Field Operator}

As a basic assumption there is postulated that the components of the tetrad field describing the gravitational
field become operators,

\begin{equation}
e^\mu_m \rightarrow \hat e^\mu_m,
\end{equation}
which fulfil the following fundamental canonical noncommutativity algebra:

\begin{equation}
\left[\hat e^\mu_m(x),\hat e^\nu_n(y)\right]=i\Lambda^{\mu\nu} \delta_{mn}\delta(x-y),
\label{commutator_tetrad}
\end{equation}
where $\Lambda^{\mu\nu}$ denotes the antisymmetric tensor describing the noncommutativity of the tetrad field and $\delta(x-y)$
denotes the delta function. This kind of quantization of the tetrad field is related to the
canonical case of noncommutativity relations between coordinates on the one hand and to the canonical quantization of a local field theory on the other hand, what becomes manifest with respect to the delta function on the left hand side of the commutation relation or quantization condition ($\ref{commutator_tetrad}$). In this sense it represents a combination of noncommutative geometry
and a certain kind of quantization of the gravitational field in the sense of a field quantization. The relation between
quantum mechanics and noncommutative geometry is analogue to the relation between canonical quantum gravity and the above
way of a quantum theoretical treatment of the tetrad field. In noncommutative geometry the commutation relation
between position and momentum, $[\hat x^i,\hat p_j]=i\delta^i_j$, is transferred to the several components of position
and leads to commutation relations between them: $[\hat x^\mu,\hat x^\nu]=i\theta^{\mu\nu}$. And in the theory based on the
algebra ($\ref{commutator_tetrad}$) the idea of a canonical quantization between a quantity describing the gravitational
field (metric, tetrad or connection) and its canonical conjugated momentum,
$[\hat h^{ab}(x),\hat p_{cd}(y)]=\frac{i}{2}\left(\delta^a_c \delta^b_d+\delta^b_c \delta^a_d\right)\delta(x-y)$
in quantum geometrodynamics for example where $h_{ab}$ describes the three metric and $p^{cd}$ its canonical conjugated
momentum, is transferred to the several components of the quantity describing the gravitational field, the tetrad field
in the consideration of this paper, leading accordingly to canonical commutation relations between them. In subsequent
sections it will become necessary to perform a series expansion of the gravitational field. To perform a series expansion
the tetrad field operator $\hat e^\mu_m$ can be expressed as the sum of the Kronecker symbol $\delta^\mu_m$ and an
operator $\hat h^\mu_m$ representing a quantum version of a small fluctuation of the gravitational field around
flat space-time,

\begin{equation}
\hat e^\mu_m=\delta^\mu_m{\bf 1}+\hat h^\mu_m,
\label{expansion}
\end{equation}
where $\hat h^\mu_m$ fulfils the same algebra as $\hat e^\mu_m$,

\begin{eqnarray}
\left[\hat e^\mu_m(x),\hat e^\nu_n(y)\right]=i\Lambda^{\mu\nu} \delta_{mn}\delta(x-y)\quad
\Leftrightarrow\quad \left[\delta^\mu_m {\bf 1}+\hat h^\mu_m (x),
\delta^\nu_n {\bf 1}+\hat h^\nu_n (y)\right]
=i\Lambda^{\mu\nu}\delta_{mn}\delta(x-y) \nonumber\\
\Leftrightarrow\quad
\left[\delta^\mu_m {\bf 1}, \delta^\nu_n {\bf 1}\right]=0\quad,\quad
\left[\delta^\mu_m {\bf 1},\hat h^\nu_n(x)\right]+\left[\hat h^\mu_m(x), \delta^\nu_n {\bf 1}\right]=0\quad,\quad
\left[\hat h^\mu_m (x),\hat h^\nu_n (y)\right]=i\Lambda^{\mu\nu}\delta_{mn}\delta(x-y).
\label{commutator_expansion}
\end{eqnarray}
This expansion will become important concerning the calculation of the generalized field equations below.
In this paper will be treated the usual case of a (3+1)-dimensional Minkowski space-time. To be able to
define coherent states for the tetrad field operator $\hat e^\mu_m$ it is of course necessary to divide
the space-time manifold into two two-dimensional submanifolds by defining the following special shape
for the noncommutativity tensor $\Lambda^{\mu\nu}$ in ($\ref{commutator_tetrad}$) according to the
consideration of the last section:

\begin{equation}
\Lambda^{\mu\nu}=\left(\begin{matrix}
0 & \lambda_a & 0 & 0\\
-\lambda_a & 0 & 0 & 0\\
0 & 0 & 0 & \lambda_b\\
0 & 0 & -\lambda_b & 0 \end{matrix}\right).
\label{tensor}
\end{equation}
To maintain Lorentz invariance there has to be chosen: $\lambda_a=\lambda_b \equiv \lambda$.
This condition of Lorentz invariance can only be fulfilled, since $\Lambda^{\mu\nu}$ is a tensor
rather than a constant matrix. The definition ($\ref{tensor}$) then implies the following
commutation relations between the components of the tetrad field being analogue to ($\ref{nc_planes}$):

\begin{equation}
\left[\hat e^1_m(x),\hat e^2_n(y)\right]=i\lambda\delta_{mn}\delta(x-y)
\quad,\quad \left[\hat e^3_m(x),\hat e^4_n(y)\right]=i\lambda\delta_{mn}\delta(x-y).
\label{commutator_components}
\end{equation}

\section{Definition of Coherent States Concerning the Gravitational Field}

In this section there will now be defined coherent states with respect to operators which are constructed
from the components of the tetrad field operator $\hat e^\mu_m$ fulfilling the special manifestation of the
algebra ($\ref{commutator_tetrad}$) induced by ($\ref{tensor}$) and leading to
($\ref{commutator_components}$). There can be followed exactly the formalism developed with respect
to noncommutative coordinates transferred to the tetrad field for the special case of a (3+1)-dimensional
Minkowski space-time. Accordingly there can be defined the following operators:

\begin{eqnarray}
\hat a_m=\frac{1}{\sqrt{2\lambda}}\left(\hat e_m^1+i\hat e_m^2\right)\quad,\quad
\hat a_m^{\dagger}=\frac{1}{\sqrt{2\lambda}}\left(\hat e_m^1-i\hat e_m^2\right),\nonumber\\
\hat b_m=\frac{1}{\sqrt{2\lambda}}\left(\hat e_m^3+i\hat e_m^4\right)\quad,\quad
\hat b_m^{\dagger}=\frac{1}{\sqrt{2\lambda}}\left(\hat e_m^3-i\hat e_m^4\right),
\label{creation_annihilation_tetrad}
\end{eqnarray}
fulfilling the commutation relations

\begin{equation}
\left[\hat a_m(x),\hat a_n^{\dagger}(y)\right]=\delta_{mn}\delta(x-y)\quad,\quad
\left[\hat b_m(x),\hat b_n^{\dagger}(y)\right]=\delta_{mn}\delta(x-y),
\label{creation_annihilation_operators}
\end{equation}
and thus behave as creation and annihilation operators. Since in the special case of $d=D/2=2$
only two pairs of creation and annihilation operators arise, they are denoted with
$a_m$,$a_m^{\dagger}$ and $b_m$, $b_m^{\dagger}$ instead of distinguishing them by an index.
The components of the tetrad field operator $\hat e^\mu_m$ can be expressed by the operators ($\ref{creation_annihilation_tetrad}$) as follows:

\begin{eqnarray}
\hat e^1_m=\sqrt{\frac{\lambda}{2}}\left(\hat a_m+\hat a_m^{\dagger}\right)\quad,\quad
\hat e^2_m=-i\sqrt{\frac{\lambda}{2}}\left(\hat a_m-\hat a_m^{\dagger}\right),\nonumber\\
\hat e^3_m=\sqrt{\frac{\lambda}{2}}\left(\hat b_m+\hat b_m^{\dagger}\right)\quad,\quad
\hat e^4_m=-i\sqrt{\frac{\lambda}{2}}\left(\hat b_m-\hat b_m^{\dagger}\right).
\end{eqnarray}
To simplify the notation the quantities $\hat E_m^\mu$ and $\hat E_m^{\mu \dagger}$ are defined 
according to

\begin{equation}
\hat E_m^\mu=\sqrt{\frac{\lambda}{2}}\left(\hat a_m,-i \hat a_m,\hat b_m,-i \hat b_m\right)\quad,\quad
\hat E_m^{\mu\dagger}=\sqrt{\frac{\lambda}{2}}\left(\hat a_m^{\dagger},i \hat a_m^{\dagger},
\hat b_m^{\dagger},i \hat b_m^{\dagger}\right),
\label{E-field}
\end{equation}
which fulfil the commutation relation

\begin{equation}
\left[\hat E_m^\mu(x),\hat E_n^{\nu \dagger}(y)\right]=i\Gamma^{\mu\nu}\delta_{mn}\delta(x-y),
\label{commutator_E-field}
\end{equation}
where the tensor $\Gamma^{\mu\nu}$ has been introduced which is defined as

\begin{equation}
\Gamma^{\mu\nu}=\frac{\lambda}{2}\left(\begin{matrix}-i&1&0&0\\-1&-i&0&0\\0&0&-i&1\\0&0&-1&-i\end{matrix}\right).
\end{equation}
The tetrad field operator $\hat e_m^\mu$ can now be expressed by using the definition ($\ref{E-field}$),

\begin{equation}
\hat e_m^\mu=\hat E_m^\mu+\hat E_m^{\mu\dagger}.
\label{tetrad_E-field_operator}
\end{equation}
Accordingly the usual tetrad field is related to the field $E_m^\mu$ and its complex conjugated field 
$E_m^{\mu *}$ according to

\begin{equation}
e_m^\mu=E_m^\mu+E_m^{\mu *}.
\label{tetrad_E-field}
\end{equation}
If there are defined coherent states within the Hilbert spaces $\mathcal{H}_a$ and $\mathcal{H}_b$ in which
the operators $\hat a_m$,$\hat a_m^{\dagger}$ respectively $\hat b_m$,$\hat b_m^{\dagger}$ act and which
represent eigenstates of $\hat a_m$ and $\hat b_m$ as ket-vectors and eigenstates of $\hat a_m^{\dagger}$
and $\hat b_m^{\dagger}$ as bra-vectors, these states of course also represent eigenstates of the
operator $\hat E_m^\mu$ and its hermitian adjoint $\hat E_m^{\mu \dagger}$. The coherent states which
depend of course on the space-time point, since they refer to field operators, are defined according to

\begin{eqnarray}
|a_m(x)\rangle=\exp\left[-\frac{|a_m(x)|^2}{2}\right]\exp\left[a_m(x)\hat a_m^{\dagger}(x)\right]|0\rangle=
\exp\left[-\frac{|a_m(x)|^2}{2}\right]\sum_{n=0}^{\infty}\frac{\left[a_m(x)\right]^n}{\sqrt{n !}}|n_{a_m}(x)\rangle, \nonumber\\
|b_m(x)\rangle=\exp\left[-\frac{|b_m(x)|^2}{2}\right]\exp\left[b_m(x)\hat b_m^{\dagger}(x)\right]|0\rangle=
\exp\left[-\frac{|b_m(x)|^2}{2}\right]\sum_{n=0}^{\infty}\frac{\left[b_m(x)\right]^n}{\sqrt{n !}}|n_{b_m}(x)\rangle,
\label{coherent_states_definition}
\end{eqnarray}
where $|n_{a_m}(x)\rangle=\frac{\left[\hat a_m^{\dagger}(x)\right]^n}{\sqrt{n !}}|0\rangle$ and
$|n_{b_m}(x)\rangle=\frac{\left[\hat b_m^{\dagger}(x)\right]^n}{\sqrt{n !}}|0\rangle$ are the eigenstates
with respect to the occupation number operators
$\hat a_m^{\dagger}(x) \hat a_m(x)$ and $\hat b_m^{\dagger}(x) \hat b_m(x)$ meaning that

\begin{equation}
\hat a_m^{\dagger}(x) \hat a_m(x)|n_{a_m}(x)\rangle=n_{a_m}(x)|n_{a_m}(x)\rangle \quad,\quad
\hat b_m^{\dagger}(x) \hat b_m(x)|n_{b_m}(x)\rangle=n_{b_m}(x)|n_{b_m}(x)\rangle,
\end{equation}
where $n_{a_m}(x)$ and $n_{b_m}(x)$ describe the corresponding eigenvalues of the occupation number operators.
The coherent states ($\ref{coherent_states_definition}$) fulfil the following eigenvalue equations:

\begin{eqnarray}
&&\hat a_m(x)|a_m(x)\rangle=a_m(x)|a_m(x)\rangle
\quad,\quad \hat b_m(x)|b_m(x)\rangle=b_m(x)|b_m(x)\rangle,\nonumber\\
&&\langle a_m(x)|\hat a_m^{\dagger}(x)=\langle a_m(x)|a_m^{*}(x)
\quad,\quad \langle b_m(x)|\hat b_m^{\dagger}(x)=\langle b_m(x)|b_m^{*}(x),
\end{eqnarray}
where $a$, $a^{*}$, $b$, $b^{*}$ are the corresponding eigenvalues to the operators
$\hat a$, $\hat a^{\dagger}$, $\hat b$, $\hat b^{\dagger}$. By building the tensor
product of the several Minkowski components of the eigenstates $|a_m \rangle$ and
$|b_m \rangle$ respectively,

\begin{equation}
|A\rangle=\prod_{m=1}^4 |a_m\rangle \quad,\quad |B\rangle=\prod_{m=1}^4 |b_m\rangle,
\end{equation}
and then building the tensor product of $|A\rangle$ and $|B\rangle$,

\begin{equation}
|\Psi\rangle=|A\rangle \otimes |B\rangle=\prod_{m=1}^4 |a_m\rangle \otimes \prod_{n=1}^4 |b_n\rangle,
\label{coherent_state_tetrad}
\end{equation}
one obtains a state $|\Psi\rangle$ being an eigenstate of all components of $\hat E_m^\mu$ meaning that

\begin{eqnarray}
\hat E_m^\mu(x) |\Psi(x)\rangle
&=&\sqrt{\frac{\lambda}{2}}\left[\hat a_m(x),-i \hat a_m(x),\hat b_m(x),-i \hat b_m(x)\right]
\prod_{n=1}^4 |a_n(x)\rangle \otimes |b_n(x)\rangle\nonumber\\
&=&\sqrt{\frac{\lambda}{2}}\left[a_m(x),-i a_m(x), b_m(x),-i b_m(x)\right]\prod_{n=1}^4
|a_n(x)\rangle \otimes |b_n(x)\rangle=E_m^\mu(x) |\Psi(x)\rangle,
\nonumber\\
\langle \Psi(x)|\hat E_m^{\mu \dagger}(x)
&=&\prod_{n=1}^4
\langle a_n(x)| \otimes \langle b_n(x)|\sqrt{\frac{\lambda}{2}}
\left[\hat a_m^{\dagger}(x),i \hat a^{\dagger}_m(x),\hat b^{\dagger}_m(x),i \hat b^{\dagger}_m(x)\right]\nonumber\\
&=&\prod_{n=1}^4 \langle a_n(x)| \otimes \langle b_n(x)|\sqrt{\frac{\lambda}{2}}
\left[a_m^{*}(x),i a_m^{*}(x), b_m^{*}(x),i b_m^{*}(x)\right]=\langle \Psi(x)|E_m^{\mu *}(x).
\label{eigenvalue_E-field}
\end{eqnarray}
This means that an expectation value can be defined for the tetrad field operator $\hat e^\mu_m$
and for all quantities depending on the tetrad field operator $\hat e^\mu_m$ by expressing the tetrad
field operator $\hat e^\mu_m$ through $\hat E^\mu_m$ and its hermitian adjoint $\hat E_m^{\mu \dagger}$
and then building expectation values with respect to $|\Psi\rangle$.
Of course $|\Psi\rangle$ has to obey the normalization condition $\langle \Psi|\Psi \rangle=1$.
According to ($\ref{expansion}$) and ($\ref{commutator_expansion}$) the operator $\hat H^\mu_m$ can
analogously be introduced which is related to $\hat h_m^\mu$ as $\hat E^\mu_m$ is related to
$\hat e^\mu_m$ according to ($\ref{tetrad_E-field_operator}$),

\begin{equation}
\hat h_m^\mu=\hat H_m^\mu+\hat H_m^{\mu \dagger},
\label{h-expansion_H-field}
\end{equation}
fulfilling the same commutation relation as $\hat E_m^\mu$ given in ($\ref{commutator_E-field}$),

\begin{equation}
\left[\hat H_m^\mu(x),\hat H_n^{\nu \dagger}(y)\right]=i\Gamma^{\mu\nu}\delta_{mn}\delta(x-y).
\label{commutator_H-field}
\end{equation}
corresponding eigenstates $|\Phi \rangle$ can be defined which correspond to the eigenstates $|\Psi\rangle$
of $\hat E_m^\mu$ for which hold the relations being analogue to ($\ref{eigenvalue_E-field}$),

\begin{eqnarray}
\hat H_m^{\mu}|\Phi \rangle=H_m^{\mu}|\Phi \rangle\quad,\quad \langle \Phi|\hat H_m^{\mu \dagger}=\langle \Phi|\hat H_m^{\mu *}.
\label{eigenstate_H}
\end{eqnarray}
The eigenstates $|\Phi \rangle$ of course have also to obey the normalization condition $\langle \Phi|\Phi \rangle=1$.

\section{Construction of Expectation Values With Respect to Coherent States}

An expectation value of any function of the tetrad field operator $f(\hat e_m^\mu)$ can be defined with respect
to a coherent state analogue to ($\ref{expectation_value_ncg}$) by expressing the tetrad field operator
$\hat e_m^\mu$ through the field operator $\hat E_m^\mu$ and its hermitian adjoint
operator $\hat E_m^{\mu \dagger}$,

\begin{equation}
\langle f(\hat e_m^\mu) \rangle=\langle \Psi|f(\hat e_m^\mu)|\Psi\rangle
=\langle \Psi|f(\hat E_m^\mu,\hat E_m^{\mu \dagger})|\Psi\rangle
=F(E_m^\mu,E_m^{\mu *})=F(e_m^\mu),
\label{expectation_value_function_tetrad}
\end{equation}
where $F(e_m^\mu)$ describes a new function depending on the usual tetrad field and generally containing additional terms. 
If the function $f(\hat e_m^\mu)=f(\hat E_m^\mu,\hat E_m^{\mu \dagger})$ depends linear on the tetrad field
operator $\hat e_m^\mu$, then the operators $\hat E_m^\mu$ and $\hat E_m^{\mu \dagger}$ can directly be
applied to $|\Psi\rangle$ and the expectation value of the function depending on the tetrad field operator $\hat e^\mu_m$
yields the function depending on the usual tetrad field $e^\mu_m$ as will be shown below using the example of the
expectation value of the single tetrad field operator $\hat e^\mu_m$ yielding the usual tetrad field $e^\mu_m$ itself.  
If $f(\hat e_m^\mu)=f(\hat E_m^\mu,\hat E_m^{\mu \dagger})$ contains products of the tetrad field operator
$\hat e_m^\mu$ and thus products of the operators $\hat E_m^\mu$ and $\hat E_m^{\mu \dagger}$, then the products
have to be permuted in such a way that all factors $\hat E_m^{\mu \dagger}$ are to the left of the
factors $\hat E_m^\mu$. This can of course be performed by using the commutation relation
($\ref{commutator_E-field}$) being a direct consequence of ($\ref{commutator_tetrad}$). 
Since there are usually just considered pointwise products of fields, the delta function within the commutator
does not appear explicitly within the calculations.
Then the eigenvalue equations ($\ref{eigenvalue_E-field}$) as well as the equation
($\ref{tetrad_E-field}$) can be used to reexpress the obtained expression in terms of the eigenvalues $E_m^\mu$
and $E_m^{\mu *}$ belonging to the operator $\hat E_m^\mu$ and its hermitian adjoint $\hat E_m^{\mu \dagger}$
respectively by the usual tetrad field $e_m^\mu$. In this way the explicit expression of the expectation value ($\ref{expectation_value_function_tetrad}$) can be calculated which yields to every function depending on the tetrad
field which is converted to a function depending on the tetrad field operator a generalized function
which depends on the usual tetrad field again. The idea is now to obtain a generalization of the dynamics
of the gravitational field according to general relativity by using expectation values of the corresponding
expressions describing the dynamics of general relativity and depending on the tetrad field operator
fulfilling the noncommutativity algebra ($\ref{commutator_tetrad}$). This means that the extended description of general
relativity according to the approach of this paper is obtained by two steps: First within the usual quantities depending
on the tetrad field the usual tetrad field $e^\mu_m$ is replaced by the tetrad field operator $\hat e^\mu_m$ defined
according to ($\ref{commutator_tetrad}$),

\begin{equation}
f(e_m^\mu) \rightarrow f(\hat e_m^\mu),
\end{equation}
and after this the expectation value between coherent states is built according to ($\ref{expectation_value_function_tetrad}$). Before approaching the generalized description of the dynamics of general relativity as it will be done in the next two sections there shall first be given some examples as a kind of preparation in this section. As already mentioned above the expectation value for the tetrad field itself corresponds to the usual tetrad field $e^\mu_m$,

\begin{eqnarray}
\langle \hat e_m^\mu \rangle=\langle \Psi|\hat e_m^\mu|\Psi \rangle
=\langle \Psi|\left(\hat E_m^\mu+\hat E_m^{\mu \dagger}\right)|\Psi \rangle
=\langle \Psi|\left(E_m^\mu+E_m^{\mu *}\right)|\Psi \rangle=E_m^\mu+E_m^{\mu *}=e_m^\mu.
\label{expectation_value_tetrad}
\end{eqnarray}
In ($\ref{expectation_value_tetrad}$) have been used ($\ref{tetrad_E-field_operator}$),($\ref{tetrad_E-field}$),
($\ref{eigenvalue_E-field}$) and of course the normalization condition $\langle \Psi|\Psi \rangle=1$. If the
expectation value of the metric field operator $\hat g^{\mu\nu}$ being related to the tetrad field
operator $\hat e^\mu_m$ according to $\hat g^{\mu\nu}=\hat e^\mu_m \hat e^\nu_n \eta^{mn}$, where $\eta^{mn}={\rm diag}\left(1,-1,-1,-1\right)$ denotes the Minkowski metric, is considered, the situation becomes already a little bit more
complicated. After replacing the tetrad field operator $\hat e^\mu_m$ by using equation ($\ref{tetrad_E-field_operator}$)
one permutation has to be performed and accordingly one obtains an additional constant term,

\begin{eqnarray}
\langle \hat g^{\mu\nu} \rangle&=&\langle \Psi|\hat g^{\mu\nu}|\Psi\rangle
=\langle \Psi| \eta^{mn}\hat e_m^\mu \hat e_n^\nu|\Psi \rangle
=\langle \Psi|\eta^{mn}\left(\hat E_m^\mu+\hat E_m^{\mu \dagger}\right)
\left(\hat E_n^\nu+\hat E_n^{\nu \dagger}\right)|\Psi \rangle\nonumber\\
&=&\langle \Psi|\eta^{mn}\left(\hat E_m^\mu \hat E_n^\nu+\hat E_m^\mu \hat E_n^{\nu \dagger}
+\hat E_m^{\mu \dagger}\hat E_n^\nu+\hat E_m^{\nu \dagger}\hat E_n^{\nu \dagger}\right)|\Psi \rangle\nonumber\\
&=&\langle \Psi|\eta^{mn}\left(\hat E_m^\mu \hat E_n^\nu+ E_n^{\nu \dagger} \hat E_m^\mu \hat
+\left[\hat E_m^\mu, E_n^{\nu \dagger}\right]
+\hat E_m^{\mu \dagger}\hat E_n^\nu+\hat E_m^{\nu \dagger}\hat E_n^{\nu \dagger}\right)|\Psi \rangle\nonumber\\
&=&\langle \Psi|\eta^{mn}\left(\hat E_m^\mu \hat E_n^\nu+\hat E_n^{\nu \dagger} \hat E_m^\mu
+\hat E_m^{\mu \dagger}\hat E_n^\nu+\hat E_m^{\nu \dagger}\hat E_n^{\nu \dagger}+i\Gamma^{\mu\nu}\delta_{mn} 
\right)|\Psi \rangle\nonumber\\
&=&\langle \Psi|\eta^{mn}\left(E_m^\mu E_n^\nu+E_n^{\nu *} E_m^\mu
+E_m^{\mu *}E_n^\nu+E_m^{\nu *}E_n^{\nu *}+i\Gamma^{\mu\nu}\delta_{mn} 
\right)|\Psi \rangle\nonumber\\
&=&\langle \Psi|\eta^{mn}e_m^\mu e_n^\nu|\Psi \rangle+\langle \Psi|i\Gamma^{\mu\nu}\eta^{mn}\delta_{mn}|\Psi \rangle
=g^{\mu\nu}-2i\Gamma^{\mu\nu}.
\end{eqnarray}
As a more complicated example this procedure of obtaining generalized functions depending on the usual tetrad field
by building of an expectation value of the usual function converted to a function depending on the tetrad field operator
shall be applied to the volume element.
Replacing the tetrad field $e^\mu_m$ by the tetrad field operator $\hat e^\mu_m$, building the expectation value
and expressing the tetrad field operator $\hat e^\mu_m$ by the operator $\hat E^\mu_m$ leads to the following
expression:

\begin{eqnarray}
\langle \hat V \rangle&=&\int d^4 x\ \langle \Psi|\det\left(\hat e_m^\mu\right)|\Psi \rangle
=\int d^4 x\ \langle \Psi|\epsilon^{abcd}\epsilon_{\mu\nu\rho\sigma}
\hat e_a^\mu \hat e_b^\nu \hat e_c^\rho \hat e_d^\sigma|\Psi \rangle\nonumber\\
&=&\int d^4 x\ \langle \Psi|\epsilon^{abcd}\epsilon_{\mu\nu\rho\sigma}
\left(\hat E_a^\mu+\hat E_a^{\mu \dagger}\right)
\left(\hat E_b^\nu+\hat E_b^{\nu \dagger}\right)
\left(\hat E_c^\rho+\hat E_c^{\rho \dagger}\right)
\left(\hat E_d^\sigma+\hat E_d^{\sigma \dagger}\right)|\Psi \rangle
\nonumber\\
&=&\int d^4 x\ \langle \Psi|\epsilon^{abcd}\epsilon_{\mu\nu\rho\sigma}
\left(\hat E_a^\mu \hat E_b^\nu \hat E_c^\rho \hat E_d^\sigma
+\hat E_a^\mu \hat E_b^\nu \hat E_c^\rho \hat E_d^{\sigma \dagger}
+\hat E_a^\mu \hat E_b^\nu \hat E_c^{\rho \dagger} \hat E_d^\sigma
+\hat E_a^\mu \hat E_b^{\nu \dagger} \hat E_c^\rho \hat E_d^\sigma
\right.\nonumber\\
&&\left. \hat E_a^{\mu \dagger} \hat E_b^\nu \hat E_c^\rho \hat E_d^\sigma
+\hat E_a^\mu \hat E_b^\nu \hat E_c^{\rho \dagger} \hat E_d^{\sigma \dagger}
+\hat E_a^\mu \hat E_b^{\nu \dagger} \hat E_c^\rho \hat E_d^{\sigma \dagger}
+\hat E_a^\mu \hat E_b^{\nu \dagger} \hat E_c^{\rho \dagger} \hat E_d^\sigma
\right.\nonumber\\
&&\left. \hat E_a^{\mu \dagger} \hat E_b^\nu \hat E_c^\rho \hat E_d^{\sigma \dagger}
+\hat E_a^{\mu \dagger} \hat E_b^\nu \hat E_c^{\rho \dagger} \hat E_d^\sigma
+\hat E_a^{\mu \dagger} \hat E_b^{\nu \dagger} \hat E_c^\rho \hat E_d^\sigma
+\hat E_a^\mu \hat E_b^{\nu \dagger} \hat E_c^{\rho \dagger} \hat E_d^{\sigma \dagger}
\right.\nonumber\\
&&\left. \hat E_a^{\mu \dagger} \hat E_b^\nu \hat E_c^{\rho \dagger} \hat E_d^{\sigma \dagger}
+\hat E_a^{\mu \dagger} \hat E_b^{\nu \dagger} \hat E_c^\rho \hat E_d^{\sigma \dagger}
+\hat E_a^{\mu \dagger} \hat E_b^{\nu \dagger} \hat E_c^{\rho \dagger} \hat E_d^\sigma
+\hat E_a^{\mu \dagger} \hat E_b^{\nu \dagger} \hat E_c^{\rho \dagger} \hat E_d^{\sigma \dagger}\right)|\Psi \rangle,
\label{volume_a}
\end{eqnarray}
where has been used that the determinant of the tetrad field operator can be expressed as follows:
$\det\left[\hat e_m^\mu\right]=\epsilon^{abcd}\epsilon_{\mu\nu\rho\sigma}\hat e_a^\mu \hat e_b^\nu
\hat e_c^\rho \hat e_d^\sigma$, where $\epsilon_{\mu\nu\rho\sigma}$ denotes the total antisymmetric
tensor in four dimensions. Permutation of the operator $\hat E_m^\mu$ and the operator
$\hat E_m^{\mu \dagger}$ by using ($\ref{commutator_E-field}$) leads to

\begin{eqnarray}
\langle \hat V \rangle&=&\int d^4 x\ \langle \Psi|\epsilon^{abcd}\epsilon_{\mu\nu\rho\sigma}
\left(\hat E_a^\mu \hat E_b^\nu \hat E_c^\rho \hat E_d^\sigma
+\hat E_a^{\sigma \dagger} \hat E_b^\mu \hat E_d^\nu \hat E_c^\rho
+i\Gamma^{\mu\sigma}\delta_{ad}\hat E_b^\nu \hat E_c^\rho
+i\hat E_a^\mu \Gamma^{\nu\sigma}\delta_{bd} \hat E_c^\rho
\right.\nonumber\\&&\left. 
+i\hat E_a^\mu \hat E_b^\nu \Gamma^{\rho\sigma}\delta_{cd}
+\hat E_c^{\rho \dagger} \hat E_a^\mu \hat E_b^\nu \hat E_d^\sigma
+i\Gamma^{\mu\rho}\delta_{ac} \hat E_b^\nu \hat E_d^\sigma
+i\hat E_a^\mu \Gamma^{\nu\rho}\delta_{bc} \hat E_d^\sigma
\right.\nonumber\\&&\left. 
+\hat E_b^{\nu \dagger} \hat E_a^\mu \hat E_c^\rho \hat E_d^\sigma
+i\Gamma^{\mu\nu}\delta_{ab} \hat E_c^\rho \hat E_d^\sigma
+\hat E_a^{\mu \dagger} \hat E_b^\nu \hat E_c^\rho \hat E_d^\sigma
+\hat E_c^{\rho \dagger} \hat E_d^{\sigma \dagger}\hat E_a^\mu \hat E_b^\nu 
\right.\nonumber\\&&\left.
+i\hat E_d^{\sigma \dagger} \hat E_a^\mu \Gamma^{\nu\rho}\delta_{bc}
-\Gamma^{\mu\sigma}\delta_{ad}\Gamma^{\nu\rho}\delta_{bc}
+i\Gamma^{\mu\rho}\delta_{ac} \hat E_d^{\sigma \dagger} \hat E_b^\nu
-\Gamma^{\mu\rho}\delta_{ac}\Gamma^{\nu\sigma}\delta_{bd}
\right.\nonumber\\&&\left.
+i\hat E_c^{\rho \dagger}\hat E_a^\mu\Gamma^{\nu\sigma}\delta_{bd}
+i\hat E_c^{\rho \dagger}\hat E_b^\nu\Gamma^{\mu\sigma}\delta_{ad}
+\hat E_b^{\nu \dagger} \hat E_d^{\sigma \dagger} \hat E_a^\mu \hat E_c^\rho 
+i\hat E_b^{\nu \dagger} \Gamma^{\mu\sigma}\delta_{ad} \hat E_c^\rho
\right.\nonumber\\&&\left.
+i\hat E_b^{\nu \dagger} \hat E_a^\mu \Gamma^{\rho\sigma}\delta_{cd}
+i\Gamma^{\mu\nu}\delta_{ab} \hat E_d^{\sigma \dagger} \hat E_c^\rho
-\Gamma^{\mu\nu}\delta_{ab} \Gamma^{\rho\sigma}\delta_{cd}
+\hat E_b^{\nu \dagger} \hat E_c^{\rho \dagger} \hat E_a^\mu \hat E_d^\sigma
\right.\nonumber\\&&\left.
+i\hat E_b^{\nu \dagger} \Gamma^{\mu\rho}\delta_{ac} \hat E_d^\sigma
+i\Gamma^{\mu\nu}\delta_{ab} \hat E_c^{\rho \dagger} \hat E_d^\sigma
+\hat E_a^{\mu \dagger} \hat E_d^{\sigma \dagger}\hat E_b^\nu \hat E_c^\rho
+i\hat E_a^{\mu \dagger} \hat E_b^\nu \Gamma^{\rho\sigma}\delta_{cd}
\right.\nonumber\\&&\left.
+i\hat E_a^{\mu \dagger} \Gamma^{\nu\sigma}\delta_{bd} \hat E_c^\rho
+\hat E_a^{\mu \dagger} \hat E_c^{\rho \dagger} \hat E_b^\nu \hat E_d^\sigma
+i\hat E_a^{\mu \dagger} \Gamma^{\nu\rho}\delta_{bc} \hat E_d^\sigma
+\hat E_a^{\mu \dagger} \hat E_b^{\nu \dagger} \hat E_c^\rho \hat E_d^\sigma
\right.\nonumber\\&&\left.
+\hat E_b^{\nu \dagger} \hat E_c^{\rho \dagger} \hat E_d^{\sigma \dagger} \hat E_a^\mu 
+i\Gamma^{\mu\nu}\delta_{ab}\hat E_c^{\rho \dagger} \hat E_d^{\sigma \dagger}
+i\hat E_b^{\nu \dagger} \Gamma^{\mu\rho}\delta_{ac} \hat E_d^{\sigma \dagger}
+i\hat E_b^{\nu \dagger} \hat E_c^{\rho \dagger} \Gamma^{\mu\sigma}\delta_{ad}
\right.\nonumber\\&&\left.
+\hat E_a^{\mu \dagger} \hat E_c^{\rho \dagger} \hat E_d^{\sigma \dagger} \hat E_b^\nu 
+i\hat E_a^{\mu \dagger} \Gamma^{\nu\rho}\delta_{bc} \hat E_d^{\sigma \dagger}
+i\hat E_a^{\mu \dagger} \hat E_c^{\rho \dagger} \Gamma^{\nu\sigma}\delta_{bd}
+\hat E_a^{\mu \dagger} \hat E_b^{\nu \dagger} \hat E_d^{\sigma \dagger} \hat E_c^\rho 
\right.\nonumber\\&&\left.
+i\hat E_a^{\mu \dagger} \hat E_b^{\nu \dagger} \Gamma^{\rho\sigma}\delta_{cd}
+\hat E_a^{\mu \dagger} \hat E_b^{\nu \dagger} \hat E_c^{\rho \dagger} \hat E_d^\sigma
+\hat E_a^{\mu \dagger} \hat E_b^{\nu \dagger} \hat E_c^{\rho \dagger} \hat E_d^{\sigma \dagger}\right)|\Psi \rangle
\label{volume_b}
\end{eqnarray}
and applying the operators to the coherent states by using ($\ref{eigenvalue_E-field}$) and then
reexpressing the obtained expression depending on the eigenvalues $E_m^\mu$ and $E_m^{\mu *}$ by the
tetrad field $e_m^\mu$ by using ($\ref{tetrad_E-field}$) finally leads to

\begin{eqnarray}
\langle \hat V \rangle&=&\int d^4 x\left[\det\left(e_m^\mu\right)+\epsilon^{abcd}\epsilon_{\mu\nu\rho\sigma}
\left(-\Gamma^{\mu\sigma}\delta_{ad}\Gamma^{\nu\rho}\delta_{bc}
-\Gamma^{\mu\rho}\delta_{ac}\Gamma^{\nu\sigma}\delta_{bd}
-\Gamma^{\mu\nu}\delta_{ab} \Gamma^{\rho\sigma}\delta_{cd}
\right.\right.\nonumber\\&&\left.\left.
+i\Gamma^{\rho\sigma}\delta_{cd} E_a^\mu E_b^\nu 
+i\Gamma^{\rho\sigma}\delta_{cd} E_a^\mu E_b^{\nu *} 
+i\Gamma^{\rho\sigma}\delta_{cd} E_a^{\mu *} E_b^\nu 
+i\Gamma^{\rho\sigma}\delta_{cd} E_a^{\mu *} E_b^{\nu *}
\right.\right.\nonumber\\&&\left.\left.
+i\Gamma^{\nu\sigma}\delta_{bd} E_a^\mu E_c^\rho
+i\Gamma^{\nu\sigma}\delta_{bd} E_a^\mu E_c^{\rho *} 
+i\Gamma^{\nu\sigma}\delta_{bd} E_a^{\mu *} E_c^\rho
+i\Gamma^{\nu\sigma}\delta_{bd} E_a^{\mu *} E_c^{\rho *}
\right.\right.\nonumber\\&&\left.\left.
+i\Gamma^{\nu\rho}\delta_{bc} E_a^\mu E_d^\sigma
+i\Gamma^{\nu\rho}\delta_{bc} E_a^\mu E_d^{\sigma *}
+i\Gamma^{\nu\rho}\delta_{bc} E_a^{\mu *} E_d^\sigma
+i\Gamma^{\nu\rho}\delta_{bc} E_a^{\mu *} E_d^{\sigma *}
\right.\right.\nonumber\\&&\left.\left.
+i\Gamma^{\mu\sigma}\delta_{ad} E_b^\nu E_c^\rho
+i\Gamma^{\mu\sigma}\delta_{ad} E_b^\nu E_c^{\rho *} 
+i\Gamma^{\mu\sigma}\delta_{ad} E_b^{\nu *}E_c^\rho
+i\Gamma^{\mu\sigma}\delta_{ad} E_b^{\nu *} E_c^{\rho *}
\right.\right.\nonumber\\&&\left.\left.
+i\Gamma^{\mu\rho}\delta_{ac} E_b^\nu E_d^\sigma
+i\Gamma^{\mu\rho}\delta_{ac} E_b^\nu E_d^{\sigma *} 
+i\Gamma^{\mu\rho}\delta_{ac} E_b^{\nu *} E_d^\sigma
+i\Gamma^{\mu\rho}\delta_{ac} E_b^{\nu *} E_d^{\sigma *}
\right.\right.\nonumber\\&&\left.\left.
+i\Gamma^{\mu\nu}\delta_{ab} E_c^\rho E_d^\sigma
+i\Gamma^{\mu\nu}\delta_{ab} E_c^\rho E_d^{\sigma *} 
+i\Gamma^{\mu\nu}\delta_{ab} E_c^{\rho *} E_d^\sigma
+i\Gamma^{\mu\nu}\delta_{ab} E_c^{\rho *} E_d^{\sigma *}
\right)\right]\nonumber\\
&=&
\int d^4 x\left[\det\left(e_m^\mu\right)+\epsilon^{abcd}\epsilon_{\mu\nu\rho\sigma}
\left(i\Gamma^{\rho\sigma}\delta_{cd} e_a^\mu e_b^\nu 
+i\Gamma^{\nu\sigma}\delta_{bd} e_a^\mu e_c^\rho
+i\Gamma^{\nu\rho}\delta_{bc} e_a^\mu e_d^\sigma
+i\Gamma^{\mu\sigma}\delta_{ad} e_b^\nu e_c^\rho
\right.\right.\nonumber\\&&\left.\left.
\quad\quad\quad+i\Gamma^{\mu\rho}\delta_{ac} e_b^\nu e_d^\sigma
+i\Gamma^{\mu\nu}\delta_{ab} e_c^\rho e_d^\sigma
-\Gamma^{\mu\sigma}\delta_{ad}\Gamma^{\nu\rho}\delta_{bc}
-\Gamma^{\mu\rho}\delta_{ac}\Gamma^{\nu\sigma}\delta_{bd}
-\Gamma^{\mu\nu}\delta_{ab} \Gamma^{\rho\sigma}\delta_{cd}
\right)\right].
\label{volume_c}
\end{eqnarray}
Thus the expectation value of the volume element operator contains many additional terms which now
also depend on the tetrad field $e^\mu_m$. In the next two sections this procedure to obtain generalized
classical expressions for the generalized quantities of general relativity formulated in terms
of the tetrad field operator $\hat e^\mu_m$ will be used to obtain generalized dynamics of general relativity.

\section{Generalized Field Equation for Matter Coupled to Gravity}

The effective field equations of matter fields as well as the gravitational field respectively in the extended
description of general relativity with components of the tetrad field which do not commute are according to the
considerations of the last section obtained by replacing the usual tetrad field $e^\mu_m$ by the tetrad field operator
$\hat e^\mu_m$ within the corresponding action, then varying the action with respect to the tetrad field operator
and finally building the expectation value of the resulting field equation containing the tetrad field operator.
Varying the action with the usual tetrad field replaced by the tetrad field operator yields of course the usual
field equation with the tetrad field replaced by the tetrad field operator. The expectation value has then to be
built according to ($\ref{expectation_value_function_tetrad}$) and yields the generalized field equation.
However, building the expectation value of the action with the usual tetrad field replaced by tetrad field operator
and then varying the resulting action which depends again on the tetrad field would lead to other field equations
which are not considered in this paper.
In this section the generalized matter field equation of a fermionic field will be considered whereas in the
next section will be considered the generalized Einstein field equation describing the dynamics of the gravitational
field itself which are both obtained by performing the procedure mentioned above.
The action of a fermionic matter field on curved space-time which is thus coupled to the gravitational field reads

\begin{equation}
S_M[e]=\int d^4 x\ \det\left[e^\mu_m\right]i\bar \psi\gamma^m
e_m^\mu\left(\partial_\mu+\frac{i}{2}\omega_\mu^{ab}[e]\Sigma_{ab}\right)\psi,
\label{usual_matter_action}
\end{equation}
where the $\gamma^m$ denote the Dirac matrices, $\bar \psi=\psi^{\dagger}\gamma^0$ and the $\Sigma_{ab}=-\frac{i}{4}\left[\gamma_a,\gamma_b\right]$ denote the generators of the Lorentz group fulfilling $\left[\Sigma_{ab},\Sigma_{cd}\right]=\eta_{bc}\Sigma_{ad}-\eta_{ac}\Sigma_{bd}+\eta_{bd}\Sigma_{ca}-\eta_{ad}\Sigma_{cb}$.
The spin connection $\omega_\mu^{ab}[e]$ depends on the tetrad field $e_\mu^m$ in the following way:

\begin{equation}
\omega_\mu^{ab}[e]=2 e^{\nu[a}\partial_{[\mu} e_{\nu]}^{b]}
+e_{\mu c}e^{\nu a}e^{\sigma b}\partial_{[\sigma}e_{\nu]}^c.
\label{spin_connection}
\end{equation}
Remark that the brackets denote antisymmetrization meaning that $[ab]=ab-ba$.
Replacing the usual tetrad field by the tetrad field operator obeying ($\ref{commutator_tetrad}$),
$e^\mu_m \rightarrow \hat e^\mu_m$, within ($\ref{usual_matter_action}$) leads to the following
action:

\begin{equation}
S_M[e]\rightarrow \hat S_M [\hat e]=\int d^4 x\ \det\left[\hat e^\mu_m\right]i\bar \psi\gamma^m 
\hat e_m^\mu\left(\partial_\mu+\frac{i}{2}\hat \omega_\mu^{ab}[\hat e]\Sigma_{ab}\right)\psi.
\label{transition_matter_action}
\end{equation}
The corresponding field equation containing the tetrad field operator $\hat e^\mu_m$ is obtained by varying
the resulting action with respect to the matter field. Building of the expectation value by using
coherent states ($\ref{coherent_state_tetrad}$) leads then to the generalized Dirac equation on curved
space-time containing the usual tetrad field $e^\mu_m$,

\begin{equation}
\langle \Psi|\frac{\delta \hat S_M[\hat e]}{\delta \bar \psi}=0|\Psi \rangle
\Leftrightarrow
\langle \Psi|\frac{1}{\det\left[\hat e_m^\mu\right]}\frac{\delta \hat S_M[\hat e]}{\delta \bar \psi}|\Psi \rangle=0.
\label{expectation_value_variation_matterfield}
\end{equation}
After concrete variation of the matter action on curved space-time depending on the tetrad field operator
$\hat e^\mu_m$ obtained in ($\ref{transition_matter_action}$) with respect to $\bar \psi$
($\ref{expectation_value_variation_matterfield}$) reads

\begin{equation}
\langle \Psi|i\gamma^m \hat e_m^\mu\left(\partial_\mu
+\frac{i}{2}\hat \omega_\mu^{ab}[\hat e]\Sigma_{ab}\right)\psi|\Psi \rangle=0,
\end{equation}
and inserting the explicit term of the spin connection ($\ref{spin_connection}$) transformed to the corresponding expression
depending on the tetrad field operator $\hat e^\mu_m$ leads to

\begin{equation}
\langle \Psi|i\gamma^m \hat e_m^\mu\left[\partial_\mu+\frac{i}{2}\left(2 \hat e^{\nu[a}\partial_{[\mu} \hat e_{\nu]}^{b]}
+\hat e_{\mu c}\hat e^{\nu a}\hat e^{\sigma b}\partial_{[\sigma}\hat e_{\nu]}^c\right)\Sigma_{ab}\right]\psi|\Psi \rangle=0.
\label{matterfield_equation_tetradoperator}
\end{equation}
To be able to treat the calculation the exact expression of ($\ref{matterfield_equation_tetradoperator}$)
will not be considered, but a series expansion of the tetrad field operator $\hat e^\mu_m$ around $\delta^\mu_m$
instead  will be considered as it has been introduced in ($\ref{expansion}$) and ($\ref{commutator_expansion}$).
As usual such a series expansion makes sense, if the perturbation $h^\mu_m$ of the classical expansion,
$e^\mu_m=\delta^\mu_m+h^\mu_m$, which becomes an operator after postulating ($\ref{commutator_tetrad}$) is assumed
to be very small. After the transition this relation looks as described by ($\ref{expansion}$),
$\hat e^\mu_m=\delta^\mu_m{\bf 1}+\hat h^\mu_m$, where $\hat h^\mu_m$ fulfils according to ($\ref{commutator_expansion}$)
the same algebra as $\hat e^\mu_m$. Concerning the further calculation there will be referred to the operators and states
defined with respect to the expansion $\hat h^\mu_m$ in
($\ref{h-expansion_H-field}$),($\ref{commutator_H-field}$),($\ref{eigenstate_H}$) and ($\ref{coherent_state_tetrad}$)
which are mathematically of course isomorphic to the operators and states defined with respect to $\hat e^\mu_m$.
In particular, a calculation to the third order in the expansion operator $\hat h^\mu_m$ will be considered, since in
case of the generalized Einstein field equation considered in the next section the terms of the first and second order do not
differ from the usual case which means that they yield no additional terms. Accordingly the expectation value of the
equation can be expressed as follows:

\begin{eqnarray}
\langle \Phi|i\gamma^m\left[\hat D_m(\hat h^0)+\hat D_m(\hat h^1)
+\hat D_m(\hat h^2)+\hat D_m(\hat h^3)\right]\psi |\Phi \rangle+\mathcal{O}\left(\hat h^4\right)=0,
\label{expectation_value_matterfield_equation_thirdorder}
\end{eqnarray}
where $\hat D_m(\hat h^0)$ describes the term of the covariant derivative which does not depend on the perturbation
of the tetrad field operator $\hat h^\mu_m$ and $\hat D_m(\hat h^1)$, $\hat D_m(\hat h^2)$ and $\hat D_m(\hat h^3)$ describe
the terms of the covariant derivative to the first, second and third order in $\hat h^\mu_m$. Accordingly $\hat D_m(\hat h^0)$,
$\hat D_m(\hat h^1)$, $\hat D_m(\hat h^2)$ and $\hat D_m(\hat h^3)$ are defined as
      
\begin{eqnarray}
\hat D_m\left(\hat h^0\right)&=&\partial_m,
\nonumber\\
\hat D_m\left(\hat h^1\right)&=&\hat h_m^\mu \partial_\mu+\frac{i}{2}\left[2 \delta^{\nu[a}\partial_{[m} \hat h_{\nu]}^{b]}+\partial^{[b}\hat h^{a]}_m\right]\Sigma_{ab},
\nonumber\\
\hat D_m\left(\hat h^2\right)&=&\frac{i}{2}\left[2 \hat h^{\nu[a}\partial_{[m} \hat h_{\nu]}^{b]}
+\hat h_{m c}\delta^{\nu a}\delta^{\sigma b}\partial_{[\sigma}\hat h_{\nu]}^c
+\delta_{m c}\hat h^{\nu a}\delta^{\sigma b}\partial_{[\sigma}\hat h_{\nu]}^c
+\delta_{m c}\delta^{\nu a}\hat h^{\sigma b}\partial_{[\sigma}\hat h_{\nu]}^c\right.\nonumber\\ &&\left.
+\hat h^\mu_m \left(2 \delta^{\nu[a}\partial_{[\mu} \hat h_{\nu]}^{b]}
+\partial^{[b}\hat h^{a]}_\mu\right)\right]\Sigma_{ab},
\nonumber\\
\hat D_m\left(\hat h^3\right)&=&\frac{i}{2}\left[
\hat h_{m c}\hat h^{\nu a}\delta^{\sigma b}\partial_{[\sigma}\hat h_{\nu]}^c
+\hat h_{m c}\delta^{\nu a}\hat h^{\sigma b}\partial_{[\sigma}\hat h_{\nu]}^c
+\delta_{m c}\hat h^{\nu a}\hat h^{\sigma b}\partial_{[\sigma}\hat h_{\nu]}^c \right. \nonumber\\ &&\left.
+\hat h^\mu_m \left(2 \hat h^{\nu[a}\partial_{[\mu} \hat h_{\nu]}^{b]}
+\hat h_{\mu c}\delta^{\nu a}\delta^{\sigma b}\partial_{[\sigma}\hat h_{\nu]}^c
+\delta_{\mu c}\hat h^{\nu a}\delta^{\sigma b}\partial_{[\sigma}\hat h_{\nu]}^c
+\delta_{\mu c}\delta^{\nu a}\hat h^{\sigma b}\partial_{[\sigma}\hat h_{\nu]}^c\right)\right]\Sigma_{ab}.
\label{Dh}
\end{eqnarray}
The expectation value of the term $i\gamma^m \hat D_m\left(\hat h^0\right)\psi$ can of course be given directly
without a long calculation, since it does not contain the operator $\hat h_m^\mu$,

\begin{eqnarray}
\langle \Phi|i\gamma^m \hat D_m\left(\hat h^0\right)\psi|\Phi \rangle
=\langle \Phi|i\gamma^m \partial_m \psi |\Phi \rangle
=i\gamma^m \partial_m \psi.
\label{expectation_value_Dh0}
\end{eqnarray}
To calculate the expectation values of the other terms it is necessary to treat the commutator of derivatives of
$\hat E_m^\mu$ or $\hat H_m^\mu$ respectively, since these terms contain such derivatives. These commutators can
be calculated as follows:

\begin{eqnarray}
\left[\partial_\rho \hat E_m^\mu(x),\partial_\sigma \hat E_n^{\nu \dagger}(y)\right]
&=&\frac{\partial}{\partial x^\rho}\frac{\partial}{\partial y^\sigma}
\left[\hat E_m^\mu(x),\hat E_n^{\nu \dagger}(y)\right]\nonumber\\
&=&\frac{\partial}{\partial x^\rho}\frac{\partial}{\partial y^\sigma}\left[i\Gamma^{\mu\nu}\delta_{mn}\delta(x-y)\right]\nonumber\\
&=&i\Gamma^{\mu\nu}\delta_{mn}\frac{\partial}{\partial x^\rho}\frac{\partial}{\partial y^\sigma}\delta(x-y)\nonumber\\
&=&-i\Gamma^{\mu\nu}\delta_{mn}\frac{\partial}{\partial x^\rho}\delta(x-y)\frac{\partial}{\partial y^\sigma}\nonumber\\
&=&i\Gamma^{\mu\nu}\delta_{mn}\delta(x-y)\frac{\partial}{\partial x^\rho}\frac{\partial}{\partial y^\sigma},
\label{commutator_derivatives_E}
\end{eqnarray}
where in the second step of ($\ref{commutator_derivatives_E}$) has been used ($\ref{commutator_E-field}$) and in the
forth and the fifth step has been used a special property of the delta function:

\begin{equation}
\int d^4 x\ f(x)\partial_\mu \delta(x-a)=-\int d^4 x\ \delta(x-a)\partial_\mu f(x).
\end{equation}
From ($\ref{commutator_derivatives_E}$) one can easily see that the following commutation relations
are valid as well:

\begin{eqnarray}
\left[\partial_\rho \hat E_m^\mu(x),\hat E_n^{\nu \dagger}(y)\right]
&=&-i\Gamma^{\mu\nu}\delta_{mn}\delta(x-y)\frac{\partial}{\partial x^\rho}\quad,\quad
\left[\hat E_m^\mu(x),\partial_\rho \hat E_n^{\nu \dagger}(y)\right]
=-i\Gamma^{\mu\nu}\delta_{mn}\delta(x-y)\frac{\partial}{\partial y^\rho},\nonumber\\
\left[\partial_\rho \partial_\sigma \hat E_m^\mu(x),\hat E_n^{\nu \dagger}(y)\right]
&=&i\Gamma^{\mu\nu}\delta_{mn}\delta(x-y)\frac{\partial}{\partial x^\rho}\frac{\partial}{\partial x^\sigma}\quad,\quad
\left[\hat E_m^\mu(x),\partial_\rho \partial_\sigma \hat E_n^{\nu \dagger}(y)\right]
=i\Gamma^{\mu\nu}\delta_{mn}\delta(x-y)\frac{\partial}{\partial y^\rho}\frac{\partial}{\partial y^\sigma}.
\label{commutator_derivatives_E_x}
\end{eqnarray}
And of course this implies that accordingly also the corresponding commutation relations containing derivatives
with respect to $H_m^\mu$ are valid,

\begin{eqnarray}
\left[\partial_\rho \hat H_m^\mu(x),\partial_\sigma \hat H_n^{\nu \dagger}(y)\right]
&=&i\Gamma^{\mu\nu}\delta_{mn}\delta(x-y)\frac{\partial}{\partial x^\rho}
\frac{\partial}{\partial y^\sigma},\nonumber\\
\left[\partial_\rho \hat H_m^\mu(x),\hat H_n^{\nu \dagger}(y)\right]&=&
-i\Gamma^{\mu\nu}\delta_{mn}\delta(x-y)\frac{\partial}{\partial x^\rho}\quad,\quad
\left[\hat H_m^\mu(x),\partial_\rho \hat H_n^{\nu \dagger}(y)\right]
=-i\Gamma^{\mu\nu}\delta_{mn}\delta(x-y)\frac{\partial}{\partial y^\rho},\nonumber\\
\left[\partial_\rho \partial_\sigma \hat H_m^\mu(x),\hat H_n^{\nu \dagger}(y)\right]
&=&i\Gamma^{\mu\nu}\delta_{mn}\delta(x-y)\frac{\partial}{\partial x^\rho}\frac{\partial}{\partial x^\sigma}\quad,\quad
\left[\hat H_m^\mu(x),\partial_\rho \partial_\sigma \hat H_n^{\nu \dagger}(y)\right]
=i\Gamma^{\mu\nu}\delta_{mn}\delta(x-y)\frac{\partial}{\partial y^\rho}
\frac{\partial}{\partial y^\sigma}.
\label{commutator_derivatives_H}
\end{eqnarray}
From ($\ref{commutator_derivatives_E}$),($\ref{commutator_derivatives_E_x}$) and ($\ref{commutator_derivatives_H}$)
it becomes clear that the commutator between derivatives of $\hat E_\mu$ and $\hat E_\mu^{\dagger}$ as well as
$\hat H_\mu$ and $\hat H_\mu^{\dagger}$ vanishes, if the expression where the commutator appears contains no further
field factors. Within the applications of the commutation relations within the calculation of the generalized actions
below the delta functions do not appear explicitly. If there appear several derivatives, they refer to the same variable,
since there always appear pointwise products of fields. This has already been the case concerning the calculation of the
volume element ($\ref{volume_a}$),($\ref{volume_b}$),($\ref{volume_c}$).
Besides the importance of the commutation relations ($\ref{commutator_derivatives_E}$),($\ref{commutator_derivatives_E_x}$)
and ($\ref{commutator_derivatives_H}$) which contain derivatives it is further decisive that the application of
derivatives of the operators $\hat E_m^\mu$ or $\hat H_m^\mu$ to a coherent state, $|\Psi \rangle$ or $|\Phi \rangle$,
yields the derivative of the corresponding eigenvalue which means that

\begin{equation}
\partial_\nu \hat E_m^\mu|\Psi\rangle=\partial_\nu E_m^\mu|\Psi\rangle
\quad,\quad \langle \Psi|\partial_\nu \hat E_m^{\mu \dagger}=\langle \Psi|\partial_\nu E_m^{\mu *}\quad,\quad
\partial_\nu \hat H_m^\mu|\Phi\rangle=\partial_\nu H_m^\mu|\Phi\rangle
\quad,\quad \langle \Phi|\partial_\nu \hat H_m^{\mu \dagger}=\langle \Phi|\partial_\nu H_m^{\mu *}.
\label{eigenvalue_equation_derivatives}
\end{equation}
The validity of the identities ($\ref{eigenvalue_equation_derivatives}$) can be shown as follows:

\begin{eqnarray}
\partial_\nu \hat E_m^\mu(x)|\Psi(x)\rangle&=&\lim_{\epsilon \to 0}
\frac{\hat E_m^\mu(x+\epsilon)-\hat E_m^\mu(x)}{\epsilon^\nu}|\Psi(x)\rangle=
\lim_{\epsilon \to 0}\frac{E_m^\mu(x+\epsilon)-E_m^\mu(x)}{\epsilon^\nu}|\Psi(x)\rangle
=\partial_\nu E_m^\mu(x)|\Psi(x)\rangle,
\end{eqnarray}
where has been used in the second step that $\lim_{\epsilon \to 0}|\Psi(x+\epsilon)\rangle=|\Psi(x)\rangle$
and therefore $\lim_{\epsilon \to 0} \hat E_m^\mu(x+\epsilon)|\Psi(x)\rangle
=\lim_{\epsilon \to 0} \hat E_m^\mu(x+\epsilon)|\Psi(x+\epsilon)\rangle
=\lim_{\epsilon \to 0} E_m^\mu(x+\epsilon)|\Psi(x+\epsilon)\rangle
=\lim_{\epsilon \to 0} E_m^\mu(x+\epsilon)|\Psi(x)\rangle$.    
To calculate the expectation values of the expressions $i\gamma^m \hat D_m\left(\hat h^1\right)\psi$,
$i\gamma^m \hat D_m\left(\hat h^2\right)\psi$ and $i\gamma^m \hat D_m\left(\hat h^3\right)\psi$
defined through ($\ref{Dh}$), the operator $\hat h_m^\mu$ within the expressions
($\ref{Dh}$) has to be replaced by $\hat H_m^\mu$ and $\hat H_m^{\mu \dagger}$ due to ($\ref{h-expansion_H-field}$),
then the permutation has to be performed by using the commutation relations ($\ref{commutator_derivatives_H}$)
and after this the operators can be applied to the coherent state $|\Phi\rangle$.
For the expression $i\gamma^m \hat D_m\left(\hat h^1\right) \psi$ defined through ($\ref{Dh}$) the commutation
relations ($\ref{commutator_derivatives_H}$) are not necessary, since there only appear linear expressions and
therefore $i\gamma^m \hat D_m\left(\hat h^1\right) \psi$ yields no additional terms compared with the classical case,

\begin{eqnarray}
\langle \Phi|i\gamma^m\hat D_m\left(\hat h^1\right) \psi|\Phi \rangle
&=&\langle \Phi|i\gamma^m\left\{\hat h^\mu_m \partial_\mu+\frac{i}{2}\left[2 \delta^{\nu[a}
\partial_{[m} \hat h_{\nu]}^{b]}+\partial^{[b}\hat h^{a]}_m\right]\Sigma_{ab}\right\}\psi|\Phi \rangle
\nonumber\\
&=&\langle \Phi|i\gamma^m\left\{\left(\hat H^\mu_m+\hat H^{\mu \dagger}_m\right)
\partial_\mu+\frac{i}{2}\left[2 \delta^{\nu[a}
\left(\partial_{[m} \hat H_{\nu]}^{b]}+\partial_{[m} \hat H_{\nu]}^{b] \dagger}\right)
+\left(\partial^{[b}\hat H^{a]}_m
+\partial^{[b}\hat H^{a] \dagger}_{m} \right)\right]\Sigma_{ab}\right\}\psi|\Phi \rangle
\nonumber\\
&=&\langle \Phi|i\gamma^m\left\{h^\mu_m \partial_\mu+\frac{i}{2}\left[2 \delta^{\nu[a}
\partial_{[m} h_{\nu]}^{b]}+\partial^{[b} h^{a]}_m \right]\Sigma_{ab}\right\}\psi|\Phi \rangle
\nonumber\\
&=&i\gamma^m\left\{h^\mu_m \partial_\mu+\frac{i}{2}\left[2 \delta^{\nu[a}
\partial_{[m} h_{\nu]}^{b]}+\partial^{[b} h^{a]}_m\right]\Sigma_{ab}\right\}\psi.
\label{expectation_value_Dh1}
\end{eqnarray}
In the second step of ($\ref{expectation_value_Dh1}$) has of course been used
($\ref{eigenvalue_equation_derivatives}$). The expectation value of $i\gamma^m D_m\left(\hat h^2\right)\psi$
defined through ($\ref{Dh}$) is calculated in the following way:

\begin{eqnarray}
\langle \Phi|i\gamma^m \hat D_m\left(\hat h^2\right)\psi|\Phi \rangle
&=&\langle \Phi|\frac{i}{2}\left[\left(2 \hat h^{\nu[a}\partial_{[m} \hat h_{\nu]}^{b]}
+\hat h_{m c}\delta^{\nu a}\delta^{\sigma b}\partial_{[\sigma}\hat h_{\nu]}^c
+\delta_{m c}\hat h^{\nu a}\delta^{\sigma b}\partial_{[\sigma}\hat h_{\nu]}^c
+\delta_{m c}\delta^{\nu a}\hat h^{\sigma b}\partial_{[\sigma}\hat h_{\nu]}^c\right)\right.\nonumber\\ &&\left.
+\hat h^\mu_m \left(2 \delta^{\nu[a}\partial_{[\mu} \hat h_{\nu]}^{b]}
+\partial^{[b}\hat h^{a]}_\mu\right)\right]i\Sigma_{ab}
\gamma^m \psi|\Phi \rangle
\nonumber\\ 
&=&\langle \Phi|\frac{i}{2}\left\{\left[2 \left(\hat H^{\nu[a}+\hat H^{\nu[a \dagger}\right)
\left(\partial_{[m} \hat H_{\nu]}^{b]}+\partial_{[m} \hat H_{\nu]}^{b]\dagger}\right)
+\left(\hat H_{m c}+\hat H_{m c}^{\dagger}\right)\delta^{\nu a}\delta^{\sigma b}
\left(\partial_{[\sigma}\hat H_{\nu]}^c+\partial_{[\sigma}\hat H_{\nu]}^{c \dagger}\right)
\right.\right.\nonumber\\ &&\left.\left.
+\delta_{m c}\left(\hat H^{\nu a}+\hat H^{\nu a \dagger}\right)\delta^{\sigma b}
\left(\partial_{[\sigma}\hat H_{\nu]}^c+\partial_{[\sigma}\hat H_{\nu]}^{c \dagger}\right)
+\delta_{m c}\delta^{\nu a}\left(\hat H^{\sigma b}+\hat H^{\sigma b \dagger}\right)
\left(\partial_{[\sigma}\hat H_{\nu]}^c+\partial_{[\sigma}\hat H_{\nu]}^{c \dagger}\right)\right]
\right.\nonumber\\ &&\left.
+\left(\hat H^\mu_m+\hat H^{\mu \dagger}_m \right) \left[2 \delta^{\nu[a}
\left(\partial_{[\mu} \hat H_{\nu]}^{b]}+\partial_{[\mu} \hat H_{\nu]}^{b] \dagger}\right)
+\left(\partial^{[b}\hat H^{a]}_\mu
+\partial^{[b}\hat H^{a]\dagger}_\mu\right)\right]\right\}i\Sigma_{ab}\gamma^m \psi|\Phi \rangle
\nonumber\\
&=&\left[2\left(h^{\nu [a}\partial_{[m} h_{\nu ]}^{b]}\chi_{ab}^m
-i\partial_{[m}\chi_{ab}^m\Gamma^\nu_{\ \nu]}\delta^{[ab]}\right)
+\delta^{\nu a}\delta^{\sigma b}\left(h_{m c}\partial_{[\sigma} h_{\nu]}^{c}\chi_{ab}^m
-4i\partial_{[\sigma}\chi_{ab}^m\Gamma_{m\nu]}\right)
\right.\nonumber\\ &&\left.
+\delta_{m c}\delta^{\sigma b}\left(h^{\nu a}\partial_{[\sigma} h_{\nu ]}^{c}\chi_{ab}^m
-i\partial_{[\sigma}\chi_{ab}^m\Gamma^\nu_{\ \nu]}\delta^{ac}\right)
+\delta_{m c}\delta^{\nu a}\left(h^{\sigma b}\partial_{[\sigma} h_{\nu ]}^{c}\chi_{ab}^m
-i\partial_{[\sigma}\chi_{ab}^m\Gamma^\sigma_{\ \nu]}\delta^{bc}\right)\right]
\nonumber\\ &&
+2\delta^{\nu[a}\left(h^\mu_m\partial_{[\mu} h_{\nu ]}^{b]}\chi_{ab}^m
-i\partial_{[\mu}\chi_{ab}^m\Gamma^\mu_{\ \nu]}\delta_m^{b]}\right)
+\left(h^\mu_m\partial^{[b} h^{a]}_\mu\chi_{ab}^m
-i\partial^{[b}\chi_{ab}^m\Gamma_{m}^{\ \ a]}\right),
\label{expectation_value_Dh2}
\end{eqnarray}
where has been defined: $\chi_{ab}^m=-\frac{1}{2}\gamma^m \Sigma_{ab}\psi$.
In the last step of ($\ref{expectation_value_Dh2}$) has been used the following identity:

\begin{eqnarray}
&&\langle \Phi|\left(\hat H_\mu^m+\hat H_\mu^{m \dagger}\right)
\left(\partial_\rho \hat H_\nu^n+\partial_\rho \hat H_\nu^{n \dagger}\right)f(x)|\Phi \rangle
\nonumber\\
&=&\langle \Phi|\left(\hat H_\mu^m \partial_\rho \hat H_\nu^n+\partial_\rho \hat H_\nu^{n \dagger} \hat H_\mu^{m}
+[\hat H_\mu^m, \partial_\rho \hat H_\nu^{n \dagger}]+\hat H_\mu^{m \dagger} \partial_\rho \hat H_\nu^n
+\hat H_\mu^{m \dagger}\partial_\rho \hat H_\nu^{n \dagger}\right)f(x)|\Phi \rangle
\nonumber\\
&=&\langle \Phi|\left(\hat H_\mu^m \partial_\rho \hat H_\nu^n+\partial_\rho \hat H_\nu^{n \dagger} \hat H_\mu^{m}
+\hat H_\mu^{m \dagger} \partial_\rho \hat H_\nu^n
+\hat H_\mu^{m \dagger}\partial_\rho \hat H_\nu^{n \dagger}-i\Gamma_{\mu\nu}\delta^{mn}\partial_\rho\right)
f(x)|\Phi \rangle
\nonumber\\
&=&\langle \Phi|\left(H_\mu^m \partial_\rho H_\nu^n+\partial_\rho H_\nu^{n *} H_\mu^{m}
+H_\mu^{m *} \partial_\rho H_\nu^n+H_\mu^{m *}\partial_\rho H_\nu^{n *}
-i\Gamma_{\mu\nu}\delta^{mn}\partial_\rho\right)f(x)|\Phi \rangle
\nonumber\\
&=&\langle \Phi|\left(H_\mu^m+H_\mu^{m *}\right)
\left(\partial_\rho H_\nu^n+\partial_\rho H_\nu^{n *}\right)f(x)
-i\Gamma_{\mu\nu}\delta^{mn}\partial_\rho f(x)|\Phi \rangle
\nonumber\\
&=&h_\mu^m \partial_\rho h_\nu^n f(x)-i\Gamma_{\mu\nu}\delta^{mn}\partial_\rho f(x),
\label{identity_quadratic}
\end{eqnarray}
where has been used ($\ref{commutator_derivatives_H}$) and $f(x)$ denotes an arbitrary field.
The expectation value of $i\gamma^m \hat D_m\left(\hat h^3\right)\psi$ defined through ($\ref{Dh}$)
can be calculated in the following way:

\begin{eqnarray}
&&\langle \Phi|i\gamma^m \hat D_m\left(\hat h^3\right)\psi|\Phi \rangle
=\langle \Phi|\frac{i}{2}\left[\hat h_{m c}\hat h^{\nu a}\delta^{\sigma b}\partial_{[\sigma}\hat h_{\nu]}^c
+\hat h_{m c}\delta^{\nu a}\hat h^{\sigma b}\partial_{[\sigma}\hat h_{\nu]}^c
+\delta_{m c}\hat h^{\nu a}\hat h^{\sigma b}\partial_{[\sigma}\hat h_{\nu]}^c \right. \nonumber\\ &&\left.
+\hat h^\mu_m \left(2 \hat h^{\nu[a}\partial_{[\mu} \hat h_{\nu]}^{b]}
+\hat h_{\mu c}\delta^{\nu a}\delta^{\sigma b}\partial_{[\sigma}\hat h_{\nu]}^c
+\delta_{\mu c}\hat h^{\nu a}\delta^{\sigma b}\partial_{[\sigma}\hat h_{\nu]}^c
+\delta_{\mu c}\delta^{\nu a}\hat h^{\sigma b}\partial_{[\sigma}\hat h_{\nu]}^c\right)\right]
i\gamma^m \Sigma_{ab}\psi|\Phi \rangle
\nonumber\\
&=&\langle \Phi|\frac{i}{2}\left\{\left(\hat H_{m c}+\hat H_{m c}^{\dagger}\right)
\left(\hat H^{\nu a}+\hat H^{\nu a \dagger}\right)\delta^{\sigma b}
\left(\partial_{[\sigma}\hat H_{\nu]}^c+\partial_{[\sigma}\hat H_{\nu]}^{c \dagger}\right)
\right.\nonumber\\ &&\left.
+\left(\hat H_{m c}+\hat H_{m c}^{\dagger}\right)\delta^{\nu a}
\left(\hat H^{\sigma b}+\hat H^{\sigma b \dagger}\right)
\left(\partial_{[\sigma}\hat H_{\nu]}^c+\partial_{[\sigma}\hat H_{\nu]}^{c \dagger} \right)
+\delta_{m c}\left(\hat H^{\nu a}+\hat H^{\nu a \dagger}\right)
\left(\hat H^{\sigma b}+\hat H^{\sigma b \dagger}\right)
\left(\partial_{[\sigma}\hat H_{\nu]}^c+\partial_{[\sigma}\hat H_{\nu]}^{c \dagger}\right)
\right. \nonumber\\ &&\left.
+\left(\hat H^\mu_m+\hat H^{\mu \dagger}_m \right)\left[2\left(\hat H^{\nu[a}+\hat H^{\nu[a \dagger}\right)
\left(\partial_{[\mu} \hat H_{\nu]}^{b]}+\partial_{[\mu} \hat H_{\nu]}^{b]\dagger}\right)
+\left(\hat H_{\mu c}+\hat H_{\mu c}^{\dagger}\right)\delta^{\nu a}\delta^{\sigma b}
\left(\partial_{[\sigma}\hat H_{\nu]}^c+\partial_{[\sigma}\hat H_{\nu]}^{c \dagger}\right)
\right.\right.\nonumber\\ &&\left.\left.
+\delta_{\mu c}\left(\hat H^{\nu a}+\hat H^{\nu a \dagger}\right)\delta^{\sigma b}
\left(\partial_{[\sigma}\hat H_{\nu]}^c+\partial_{[\sigma}\hat H_{\nu]}^{c \dagger}\right)
+\delta_{\mu c}\delta^{\nu a}\left(\hat H^{\sigma b}+\hat H^{\sigma b \dagger}\right)
\left(\partial_{[\sigma}\hat H_{\nu]}^c+\partial_{[\sigma}\hat H_{\nu]}^{c \dagger}
\right)\right]\right\}i\gamma^m\Sigma_{ab}\psi|\Phi \rangle
\nonumber\\
&=&\delta^{\sigma b}\left[
h_{m c} h^{\nu a} \partial_\sigma h_\nu^c \chi_{ab}^m
-i\Gamma_{m}^{\ \nu}\partial_\sigma h_\nu^a \chi_{ab}^m
-4 i\Gamma_{m\nu}\partial_\sigma \left(h^{\nu a} \chi_{ab}^m\right)
-i\Gamma^\nu_{\ \nu} \partial_\sigma \left(h_{m}^{a} \chi_{ab}^m\right)\right]
\nonumber\\ &&
+\delta^{\nu a}\left[
h_{m c} h^{\sigma b} \partial_\sigma h_\nu^c \chi_{ab}^m
-i\Gamma_{m}^{\ \sigma}\partial_\sigma h_\nu^b \chi_{ab}^m
-4 i\Gamma_{m\nu}\partial_\sigma \left(h^{\sigma b} \chi_{ab}^m\right)
-i\Gamma^\sigma_{\ \nu}\partial_\sigma \left(h_m^b \chi_{ab}^m\right)\right]
\nonumber\\ &&
+\delta_{m c}\left[
h^{\nu a} h^{\sigma b} \partial_\sigma h_\nu^c \chi_{ab}^m
-i\Gamma^{\nu\sigma}\delta^{ab} \partial_\sigma h_\nu^c \chi_{ab}^m
-i\Gamma^\nu_{\ \nu}\delta^{ac} \partial_\sigma \left(h^{\sigma b} \chi_{ab}^m\right)
-i\Gamma^\sigma_{\ \nu}\delta^{bc} \partial_\sigma \left(h^{\nu a} \chi_{ab}^m\right)\right]
\nonumber\\ &&
+2\left[h^\mu_m h^{\nu[ a} \partial_{[\mu} h_{\nu]}^{b]} \chi_{ab}^m
-i\Gamma^{\mu\nu}\delta_m^{[a} \partial_{[\mu} h_{\nu]}^{b]} \chi_{ab}^m
-i\partial_{[\mu} \left(h^{\nu [a} \chi_{ab}^m\right)\Gamma^\mu_{\ \nu]}\delta_m^{b]}
-i\partial_{[\mu} \left(h^\mu_m \chi_{ab}^m\right)\Gamma^{\nu}_{\ \nu]}\delta^{[ab]}\right]
\nonumber\\ &&
+\delta^{\nu a}\delta^{\sigma b}\left[h^\mu_m h_{\mu c} \partial_{[\sigma} h_{\nu]}^{c} \chi_{ab}^m
-i\Gamma^{\mu}_{\mu}\delta_{mc} \partial_{[\sigma} h_{\nu]}^{c} \chi_{ab}^m
-i\partial_{[\sigma} \left(h_{\mu c} \chi_{ab}^c\right)\Gamma^\mu_{\ \nu]}
-4 i\partial_{[\sigma} \left(h^\mu_m \chi_{ab}^m\right)\Gamma_{\mu\nu]}\right]
\nonumber\\ &&
+\delta_{\mu c}\delta^{\sigma b}\left[h^\mu_m h^{\nu a} \partial_{[\sigma} h_{\nu]}^{c} \chi_{ab}^m
-i\Gamma^{\mu\nu} \partial_{[\sigma} h_{\nu]}^{c} \chi_{ab}^a
-i\partial_{[\sigma} \left(h^{\nu a} \chi_{ab}^c\right)\Gamma^\mu_{\ \nu]}
-i\partial_{[\sigma} \left(h^\mu_m \chi_{ab}^m\right)\Gamma^{\nu}_{\ \nu]}\delta^{ac}\right]
\nonumber\\ &&
+\delta_{\mu c}\delta^{\nu a}\left[h^\mu_m h^{\sigma b} \partial_{[\sigma} h_{\nu]}^{c} \chi_{ab}^m
-i\Gamma^{\mu\sigma}\partial_{[\sigma} h_{\nu]}^{c} \chi_{ab}^b
-i\partial_{[\sigma} \left(h^{\sigma b} \chi_{ab}^c\right)\Gamma^\mu_{\ \nu]}
-i\partial_{[\sigma} \left(h^\mu_m \chi_{ab}^m\right)\Gamma^{\sigma}_{\ \nu]}\delta^{bc}\right].
\label{expectation_value_Dh3}
\end{eqnarray}
In the last step of ($\ref{expectation_value_Dh3}$) has been used the following identity:

\begin{eqnarray}
&&\langle \Phi|\left(\hat H_\mu^a+\hat H_\mu^{a \dagger}\right)
\left(\hat H_\nu^b+\hat H_\nu^{b \dagger}\right)\left(\partial_\sigma \hat H_\rho^{c}
+\partial_\sigma \hat H_\rho^{c \dagger}\right)f(x)|\Phi \rangle
\nonumber\\
&=&\langle \Phi|\left[\hat H_\mu^a \hat H_\nu^b \partial_\sigma \hat H_\rho^{c}
+\hat H_\mu^a \hat H_\nu^b \partial_\sigma \hat H_\rho^{c \dagger}
+\hat H_\mu^a, \hat H_\nu^{b \dagger} \partial_\sigma \hat H_\rho^{c}
+\hat H_\mu^a \hat H_\nu^{b \dagger} \partial_\sigma \hat H_\rho^{c \dagger}
\right.\nonumber\\&&\left.
+\hat H_\mu^{a \dagger} \hat H_\nu^b \partial_\sigma \hat H_\rho^{c}
+\hat H_\mu^{a \dagger} \hat H_\nu^b \partial_\sigma \hat H_\rho^{c \dagger}
+\hat H_\mu^{a \dagger} \hat H_\nu^{b \dagger} \partial_\sigma \hat H_\rho^{c}
+\hat H_\mu^{a \dagger} \hat H_\nu^{b \dagger} \partial_\sigma \hat H_\rho^{c \dagger}\right]f(x)|\Psi \rangle
\nonumber\\
&=&\langle \Psi|\left[\hat H_\mu^a \hat H_\nu^b \partial_\sigma \hat H_\rho^{c}
+\partial_\sigma \hat H_\rho^{c \dagger} \hat H_\mu^a \hat H_\nu^b 
+\left[\hat H_\mu^a, \partial_\sigma \hat H_\rho^{c \dagger}\right] \hat H_\nu^b 
+\hat H_\mu^a \left[\hat H_\nu^b, \partial_\sigma \hat H_\rho^{c \dagger}\right]
\right.\nonumber\\&&\left.
+\hat H_\nu^{b \dagger} \hat H_\mu^a  \partial_\sigma \hat H_\rho^{c}
+\left[\hat H_\mu^a, \hat H_\nu^{b \dagger}\right] \partial_\rho \hat H_\rho^{c}
+\hat H_\nu^{b \dagger} \hat \partial_\sigma \hat H_\rho^{c \dagger} H_\mu^a
+\hat H_\nu^{b \dagger}\left[\hat H_\mu^a, \partial_\sigma \hat H_\rho^{c \dagger}\right]
+\left[\hat H_\mu^a, \hat H_\nu^{b \dagger}\right]\partial_\sigma \hat H_\rho^{c \dagger}
\right.\nonumber\\&&\left.
+\hat H_\mu^{a \dagger} \hat H_\nu^b \partial_\sigma \hat H_\rho^{c}
+\hat H_\mu^{a \dagger} \partial_\sigma \hat H_\rho^{c \dagger} \hat H_\nu^b 
+\hat H_\mu^{a \dagger} \left[\hat H_\nu^b, \partial_\sigma \hat H_\rho^{c \dagger}\right]
+\hat H_\mu^{a \dagger} \hat H_\nu^{b \dagger} \partial_\sigma \hat H_\rho^{c}
+\hat H_\mu^{a \dagger} \hat H_\nu^{b \dagger} \partial_\sigma \hat H_\rho^{c \dagger}\right]f(x)|\Phi \rangle
\nonumber\\
&=&\langle \Phi|\left[\hat H_\mu^a \hat H_\nu^b \partial_\sigma \hat H_\rho^{c} f(x)
+\partial_\sigma \hat H_\rho^{c \dagger} \hat H_\mu^a \hat H_\nu^b f(x)
-i\Gamma_{\mu\rho}\delta^{ac} \partial_\sigma \left(\hat H_\nu^b f(x)\right)
-i\Gamma_{\nu\rho}\delta^{bc} \partial_\sigma \left(\hat H_\mu^a f(x)\right) 
\right.\nonumber\\&&\left.
+\hat H_\nu^{b \dagger} \hat H_\mu^a  \partial_\sigma \hat H_\rho^{c} f(x)
-i\Gamma_{\mu\nu}\delta^{ab} \partial_\sigma \hat H_\rho^{c} f(x)
+\hat H_\nu^{b \dagger} \hat \partial_\sigma \hat H_\rho^{c \dagger} H_\mu^a f(x)
-i\Gamma_{\mu\rho}\delta^{ac} \partial_\sigma \left(\hat H_\nu^{b \dagger} f(x)\right)
-i\Gamma_{\mu\nu}\delta^{ab} \partial_\sigma \hat H_\rho^{c \dagger} f(x)
\right.\nonumber\\&&\left.
+\hat H_\mu^{a \dagger} \hat H_\nu^b \partial_\sigma \hat H_\rho^{c} f(x)
+\hat H_\mu^{a \dagger} \partial_\sigma \hat H_\rho^{c \dagger} \hat H_\nu^b f(x) 
-i\Gamma_{\nu\rho}\delta^{bc} \partial_\sigma \left(\hat H_\mu^{a \dagger}f(x)\right)
+\hat H_\mu^{a \dagger} \hat H_\nu^{b \dagger} \partial_\sigma \hat H_\rho^{c} f(x)
+\hat H_\mu^{a \dagger} \hat H_\nu^{b \dagger} \partial_\sigma \hat H_\rho^{c \dagger} f(x)\right]|\Phi \rangle
\nonumber\\
&=&\langle \Phi|\left\{H_\mu^a H_\nu^b \partial_\sigma H_\rho^{c}f(x)
+\partial_\sigma H_\rho^{c *} H_\mu^a H_\nu^b f(x)
+H_\nu^{b *} H_\mu^a  \partial_\sigma H_\rho^{c} f(x)
+H_\nu^{b *} \partial_\sigma H_\rho^{c *} H_\mu^a f(x)
\right.\nonumber\\&&\left.
+H_\mu^{a *} H_\nu^b \partial_\sigma H_\rho^{c} f(x)
+H_\mu^{a *} \partial_\sigma H_\rho^{c *} H_\nu^b f(x)
+H_\mu^{a *} H_\nu^{b *} \partial_\sigma H_\rho^{c} f(x)
+H_\mu^{a *} H_\nu^{b *} \partial_\sigma H_\rho^{c *} f(x)
\right.\nonumber\\&&\left.
-i\Gamma_{\mu\nu}\delta^{ab} \left[\partial_\sigma H_\rho^{c}
+\partial_\sigma H_\rho^{c *}\right] f(x)
-i\Gamma_{\mu\rho}\delta^{ac} \left[\partial_\sigma \left(H_\nu^b f(x)\right)
+\partial_\sigma \left(H_\nu^{b *} f(x)\right)\right]
\right.\nonumber\\&&\left.
-i\Gamma_{\nu\rho}\delta^{bc} \left[\partial_\sigma \left(H_\mu^a f(x)\right) 
+\partial_\sigma \left(H_\mu^{a *}f(x)\right)\right]\right\}|\Phi \rangle
\nonumber\\
&=&h_\mu^a h_\nu^b \partial_\sigma h_\rho^c f(x)
-i\Gamma_{\mu\nu}\delta^{ab} \partial_\sigma h_\rho^c f(x)
-i\Gamma_{\mu\rho}\delta^{ac} \partial_\sigma \left[h_\nu^b f(x)\right]
-i\Gamma_{\nu\rho}\delta^{bc} \partial_\sigma \left[h_\mu^a f(x)\right],
\label{identity_third_power}
\end{eqnarray}
where has been used ($\ref{commutator_derivatives_H}$) again and from which can be seen directly that it also holds

\begin{eqnarray}
&&\langle \Phi|\left(\hat H_\mu^a+\hat H_\mu^{a \dagger}\right)
\left(\partial_\sigma \hat H_\nu^b+\partial_\sigma \hat H_\nu^{b \dagger}\right)\left(\partial_\lambda \hat H_\rho^{c}
+\partial_\lambda \hat H_\rho^{c \dagger}\right)f(x)|\Phi \rangle \nonumber\\
&=&h_\mu^a \partial_\sigma h_\nu^b \partial_\lambda h_\rho^c f(x)
+i\Gamma_{\mu\nu}\delta^{ab} \partial_\sigma \left[\partial_\lambda h_\rho^c f(x) \right]
+i\Gamma_{\mu\rho}\delta^{ac} \partial_\lambda \left[\partial_\sigma h_\nu^b f(x) \right]
+i\Gamma_{\nu\rho}\delta^{bc} \partial_\sigma \partial_\lambda \left[h_\mu^a f(x) \right],    
\label{identity_third_power_second}
\end{eqnarray}
what will be become important concerning the derivation of the generalized Einstein field equation in the next section.
Thus the generalized matter field equation to the third order in $h^\mu_m$ is given by ($\ref{expectation_value_matterfield_equation_thirdorder}$) with the concrete expressions for the expectation
values calculated in ($\ref{expectation_value_Dh0}$),($\ref{expectation_value_Dh1}$),($\ref{expectation_value_Dh2}$)
and ($\ref{expectation_value_Dh3}$).

\section{Generalized Einstein Field Equation}

In the last section has been derived the expectation value of the matter field equation referring to a fermionic field
containing the tetrad field operator $\hat e^\mu_m$ which is nothing else than a generalized Dirac equation on curved
space-time. In this section there will be considered the generalized dynamics of the gravitational field itself which is
usually described by the Einstein field equation. In accordance with the above derivation of the generalized matter field
equation the Einstein-Hilbert action with the tetrad field replaced by the tetrad field operator has to be considered 
to obtain the correct generalized Einstein field equation. The usual Einstein-Hilbert action expressed by the tetrad field $e^\mu_m$ reads

\begin{equation}
S_{EH}[e]=\int d^4 x\ \det\left[e_m^\mu\right]e^\mu_a e^\nu_b R_{\mu\nu}^{ab}[e],
\label{EH_action}
\end{equation}
where the Riemann tensor $R_{\mu\nu}^{ab}$ is expressed by the spin connection $\omega_\nu^{ab}$ as follows:

\begin{equation}
R_{\mu\nu}^{ab}[e]=\partial_\mu \omega_\nu^{ab}[e]-\partial_\nu \omega_\mu^{ab}[e]
+\omega_\mu^{ac}[e]\omega_\nu^{cb}[e]-\omega_\nu^{ac}[e]\omega_\mu^{cb}[e],
\label{Riemann_tensor}
\end{equation}
the spin connection $\omega_\mu^{ab}[e]$ depends on the tetrad field $e^\mu_m$ according to ($\ref{spin_connection}$)
and therefore inserting the corresponding expression for the spin connection ($\ref{spin_connection}$)
to the Riemann tensor ($\ref{Riemann_tensor}$) yields the following expression for the Riemann tensor in dependence
on the tetrad field $e^\mu_m$:

\begin{eqnarray}
R_{\mu\nu}^{ab}&=&\partial_\mu \left(2 e^{\rho[a}\partial_{[\nu} e_{\rho]}^{b]}
+e_{\nu d}e^{\rho a}e^{\sigma b}\partial_{[\sigma}e_{\rho]}^d \right)
-\partial_\nu \left(2 e^{\rho[a}\partial_{[\mu} e_{\rho]}^{b]}
+e_{\mu d}e^{\rho a}e^{\sigma b}\partial_{[\sigma}e_{\rho]}^d \right)
\\
&&+\left(4 e^{\rho[a}\partial_{[\mu} e_{\rho]}^{c]} e^{\lambda[c}\partial_{[\nu} e_{\lambda]}^{b]}
+2 e^{\rho[a}\partial_{[\mu} e_{\rho]}^{c]} e_{\nu e}e^{\lambda c} e^{\kappa b}\partial_{[\kappa}e_{\lambda]}^e
+2 e_{\mu d}e^{\rho a}e^{\sigma c}\partial_{[\sigma}e_{\rho]}^d e^{\lambda[c}\partial_{[\nu} e_{\lambda]}^{b]}
+e_{\mu d}e^{\rho a}e^{\sigma c}\partial_{[\sigma}e_{\rho]}^d e_{\nu e}e^{\lambda c}e^{\kappa b}
\partial_{[\kappa}e_{\lambda]}^e\right)
\nonumber\\
&&-\left(4 e^{\rho[a}\partial_{[\nu} e_{\rho]}^{c]} e^{\lambda[c}\partial_{[\mu} e_{\lambda]}^{b]}
+2 e^{\rho[a}\partial_{[\nu} e_{\rho]}^{c]} e_{\mu e}e^{\lambda c}e^{\kappa b}\partial_{[\kappa}e_{\lambda]}^e
+2 e_{\nu d}e^{\rho a}e^{\sigma c}\partial_{[\sigma}e_{\rho]}^d e^{\lambda[c}\partial_{[\mu} e_{\lambda]}^{b]}
+e_{\nu d}e^{\rho a}e^{\sigma c}\partial_{[\sigma}e_{\rho]}^d e_{\mu e}e^{\lambda c}e^{\kappa b}
\partial_{[\kappa}e_{\lambda]}^e\right).\nonumber
\end{eqnarray}
To obtain the dynamics of the generalized theory analogue to the matter action the transition $e^\mu_m \rightarrow \hat e^\mu_m$ has to be performed with respect to the Einstein-Hilbert action ($\ref{EH_action}$),

\begin{equation}
S_{EH}[e]\rightarrow \hat S_{EH}[\hat e]=\int d^4 x
\ \det\left[\hat e_m^\mu\right]\hat e^\mu_a \hat e^\nu_b \hat R_{\mu\nu}^{ab}[\hat e].
\end{equation}
The obtained action $\hat S_{EH}[\hat e]$ has to be varied with respect to $\hat e^\mu_m$ to derive
the corresponding Einstein field equation containing the tetrad field operator $\hat e^\mu_m$. The
generalized Einstein field equation containing the usual tetrad field $e^\mu_m$ can then be inferred by
building the expectation value of this equation in complete analogy to the case of the derivation of
the generalized matter field equation,

\begin{equation}
\langle \Psi|\frac{\delta \hat S_{EH}[\hat e]}{\delta \hat e_a^\mu}=0|\Psi\rangle
\Leftrightarrow
\langle \Psi|\frac{1}{\det\left[\hat e_a^\mu\right]}\frac{\delta \hat S_{EH}[\hat e]}{\delta \hat e_a^\mu}|\Psi\rangle=0
\Leftrightarrow \langle \Psi|\hat G_\mu^a[\hat e]|\Psi\rangle
=\langle \Psi|\hat R_\mu^a[\hat e]-\frac{1}{2}\hat R[\hat e] \hat e_\mu^a|\Psi \rangle=0,
\label{expectation_value_Einstein_free}
\end{equation}
where has been used the definition of the generalized Einstein tensor:
$\hat G_\mu^a[\hat e]=\hat R_\mu^a[\hat e]-\frac{1}{2}\hat R[\hat e] \hat e_\mu^a$.
($\ref{expectation_value_Einstein_free}$) describes the expectation value of the Einstein field equation of
the free gravitational field expressed in terms of the tetrad field operator $\hat e^\mu_m$. To obtain the
expectation value of the Einstein field equation containing the tetrad field operator in the presence of
matter the generalized matter action obtained in ($\ref{transition_matter_action}$) for example has to be
included leading to
  
\begin{eqnarray}
&&\langle \Psi|\frac{1}{\det[\hat e^\mu_a]}\left(\frac{\delta \hat S_{EH}[\hat e]}{\delta \hat e^\mu_a}+
\frac{\delta \hat S_M[\hat e]}{\delta \hat e^\mu_a}\right)|\Psi \rangle=0
\Leftrightarrow
\langle \Psi|\hat e^\nu_b \hat R_{\mu\nu}^{ab}[\hat e]
-\frac{1}{2}\hat e^\nu_b \hat e^\rho_c \hat R_{\nu\rho}^{bc}[\hat e]\hat e_\mu^a|\Psi \rangle
=-8 \pi G\langle \Psi|\hat T_\mu^a[\hat e]|\Psi \rangle \nonumber\\
&&\Leftrightarrow
\langle \Psi|\hat G_\mu^a[\hat e]|\Psi\rangle
=\langle \Psi|\hat R_\mu^a[\hat e]-\frac{1}{2}\hat R[\hat e] \hat e_\mu^a|\Psi \rangle
=-8 \pi G\langle \Psi|\hat T_\mu^a[\hat e]|\Psi \rangle,
\label{expectation_value_Einstein_matter}
\end{eqnarray}
where have been used the definitions of the generalized Ricci tensor, $\hat R_{\mu}^{a}=\hat e^\nu_b \hat R_{\mu\nu}^{ab}$,
and the generalized Ricci scalar, $\hat R=\hat e^{\mu}_a \hat e^{\nu}_b \hat R_{\mu\nu}^{ab}$, as well as the definition of the
energy momentum tensor transferred to the definition of the generalized energy momentum tensor
depending on the tetrad field operator $\hat e^\mu_m$,

\begin{equation}
\hat T_\mu^a[\hat e]=\frac{1}{\det[\hat e^\mu_a]}\frac{\delta \hat S_M[\hat e]}{\delta \hat e_a^\mu},
\end{equation}
where the matter action $\hat S_M[\hat e]$ depending on the tetrad field operator $\hat e^\mu_m$ is defined according to ($\ref{transition_matter_action}$) for example, if the matter field is a fermionic field.
To calculate the concrete expression of the expectation value of the generalized Einstein field equation
($\ref{expectation_value_Einstein_matter}$), again a series expansion according to ($\ref{expansion}$) has to
be performed, $\hat e^\mu_m=\delta^\mu_m{\bf 1}+\hat h^\mu_m$, leading to an expectation value of the
Einstein field equation containing the tetrad field operator which is of the following shape:

\begin{equation}
\langle \Phi|\hat G_\mu^a[\hat h^1]+\hat G_\mu^a[\hat h^2]+\hat G_\mu^a[\hat h^3]|\Phi \rangle+\mathcal{O}\left(\hat h^4\right)
=-8 \pi G\langle \Phi|\hat T_\mu^a[\hat h,\hat h^2,\hat h^3]|\Phi \rangle+\mathcal{O}\left(\hat h^4\right),
\label{expectation_value_Einstein_matter_thirdorder}
\end{equation}
where $\hat G_\mu^a[\hat h^1]$, $\hat G_\mu^a[\hat h^2]$ and $\hat G_\mu^a[\hat h^3]$ denote the expressions of the
Einstein tensor depending linear, quadratic and to the third power on $\hat h^\mu_m$ and have the following shape,
where $\mu \leftrightarrow \nu$ denotes the term in the bracket with the indices $\mu$ and $\nu$ exchanged:

\begin{eqnarray}
\hat G_\mu^a[\hat h^1]&=&\left[\partial_\mu \left(2 \delta^{\rho[a}\partial_{[b} \hat h_{\rho]}^{b]}
+\partial^{[b}\hat h^{a]}_b \right)
-\mu \leftrightarrow b \right]
-\frac{1}{2} \left[
\partial_b \left(2 \delta^{\sigma[b}\partial_{[c} \hat h_{\sigma]}^{c]}
+\partial^{[c}\hat h^{b]}_c \right)
-b \leftrightarrow c \right]\delta_\mu^a,
\label{Gterm_h1}
\end{eqnarray}

\begin{eqnarray}
&&\hat G_\mu^a[\hat h^2]=\left[\partial_\mu \left(2 \hat h^{\rho[a}\partial_{[b} \hat h_{\rho]}^{b]}
+\hat h_{b d}\delta^{\rho a}\delta^{\sigma b}\partial_{[\sigma}\hat h_{\rho]}^d
+\delta_{b d}\hat h^{\rho a}\delta^{\sigma b}\partial_{[\sigma}\hat h_{\rho]}^d
+\delta_{b d}\delta^{\rho a}\hat h^{\sigma b}\partial_{[\sigma}\hat h_{\rho]}^d \right)
\right.\nonumber
\\&&\left.+4 \delta^{\rho[a}\partial_{[\mu} \hat h_{\rho]}^{c]}
\delta^{\lambda[c}\partial_{[b} \hat h_{\lambda]}^{b]}
+2 \delta^{\rho[a}\partial_{[\mu} \hat h_{\rho]}^{c]}\partial^{[b} \hat h^{c]}_b
+2\partial^{[c} \hat h^{a]}_\mu
\delta^{\lambda[c}\partial_{[b} \hat h_{\lambda]}^{b]}
+\partial^{[c} \hat h^{a]}_\mu \partial^{[b} \hat h^{c]}_b
-\mu \leftrightarrow b \right]
\nonumber\\&&
+\hat h^\nu_b \left[\partial_\mu \left(2 \delta^{\rho[a}\partial_{[\nu} \hat h_{\rho]}^{b]}
+\partial^{[b}\hat h^{a]}_\nu\right)
-\mu \leftrightarrow \nu \right]
\nonumber\\
&&-\frac{1}{2} \left[
\partial_b \left(2 \hat h^{\lambda[b}\partial_{[c} \hat h_{\lambda]}^{c]}
+\hat h_{c e} \delta^{\lambda b} \delta^{\sigma c}\partial_{[\sigma}\hat h_{\lambda]}^e
+\delta_{c e} \hat h^{\lambda b} \delta^{\sigma c}\partial_{[\sigma}\hat h_{\lambda]}^e
+\delta_{c e} \delta^{\lambda b} \hat h^{\sigma c}\partial_{[\sigma}\hat h_{\lambda]}^e \right)
\right. \nonumber\\&&\left.
+4 \delta^{\tau[b}\partial_{[b} \hat h_{\tau]}^{d]}
\delta^{\lambda[d}\partial_{[c} \hat h_{\lambda]}^{c]}
+2 \delta^{\tau[b}\partial_{[b} \hat h_{\tau]}^{d]}\partial^{[c}\hat h^{d]}_c
+2 \partial^{[d}\hat h^{b]}_b
\delta^{\lambda[d}\partial_{[c} \hat h_{\lambda]}^{c]}
+\partial^{[d}\hat h^{b]}_b \partial^{[c}\hat h^{d]}_c
-b \leftrightarrow c \right] \delta_\mu^a
\nonumber\\&&
-\frac{1}{2}\hat h_b^\nu \left[
\partial_\nu \left(2 \delta^{\sigma[b}\partial_{[c} \hat h_{\sigma]}^{c]}
+\partial^{[c}\hat h^{b]}_c \right)
-\nu \leftrightarrow c
\right] \delta_\mu^a
-\frac{1}{2}\hat h_c^\rho\left[
\partial_b \left(2 \delta^{\sigma[b}\partial_{[\rho} \hat h_{\sigma]}^{c]}
+\partial^{[c}\hat h^{b]}_\rho \right)
-b \leftrightarrow \rho
\right] \delta_\mu^a
\nonumber\\&&
-\frac{1}{2} \left[
\partial_b\left(2 \delta^{\sigma[b}\partial_{[c} \hat h_{\sigma]}^{c]}
+\partial^{[c}\hat h^{b]}_c \right)
-b \leftrightarrow c
\right]\hat h_\mu^a,
\label{Gterm_h2}
\end{eqnarray}

\begin{eqnarray}
&&\hat G_\mu^a[\hat h^3]
=\left[\partial_\mu \left(
\hat h_{b d}\hat h^{\rho a}\delta^{\sigma b}\partial_{[\sigma}\hat h_{\rho]}^d 
+\hat h_{b d}\delta^{\rho a}\hat h^{\sigma b}\partial_{[\sigma}\hat h_{\rho]}^d 
+\delta_{b d}\hat h^{\rho a}\hat h^{\sigma b}\partial_{[\sigma}\hat h_{\rho]}^d \right)
+4\hat h^{\rho[a}\partial_{[\mu} \hat h_{\rho]}^{c]} \delta^{\lambda[c}\partial_{[b} \hat h_{\lambda]}^{b]}
\right.\nonumber\\&&\left.
+2\hat h^{\rho[a}\partial_{[\mu} \hat h_{\rho]}^{c]}\partial^{[b} \hat h^{c]}_b
+2\hat h_{\mu d}\delta^{\rho a}\delta^{\sigma c}\partial_{[\sigma} \hat h_{\rho]}^d
\delta^{\lambda[c}\partial_{[b} \hat h_{\lambda]}^{b]}
+2\delta_{\mu d}\hat h^{\rho a}\delta^{\sigma c}\partial_{[\sigma} \hat h_{\rho]}^d
\delta^{\lambda[c}\partial_{[b} \hat h_{\lambda]}^{b]}
+2\delta_{\mu d}\delta^{\rho a}\hat h^{\sigma c}\partial_{[\sigma} \hat h_{\rho]}^d
\delta^{\lambda[c}\partial_{[b} \hat h_{\lambda]}^{b]}
\right.\nonumber\\&&\left.
+\hat h_{\mu d}\delta^{\rho a}\delta^{\sigma c}\partial_{[\sigma} \hat h_{\rho]}^d \partial^{[b} \hat h^{c]}_b
+\delta_{\mu d}\hat h^{\rho a}\delta^{\sigma c}\partial_{[\sigma} \hat h_{\rho]}^d\partial^{[b} \hat h^{c]}_b
+\delta_{\mu d}\delta^{\rho a}\hat h^{\sigma c}\partial_{[\sigma} \hat h_{\rho]}^d\partial^{[b} \hat h^{c]}_b
+4\delta^{\rho[a}\partial_{[\mu} \hat h_{\rho]}^{c]}
\hat h^{\lambda[c}\partial_{[b} \hat h_{\lambda]}^{b]}
\right.\nonumber\\&&\left.
+2\partial^{[c} \hat h^{a]}_\mu\hat h^{\lambda[c}\partial_{[b} \hat h_{\lambda]}^{b]}
+2\delta^{\rho[a}\partial_{[\mu} \hat h_{\rho]}^{c]}
\hat h_{b e}\delta^{\lambda c}\delta^{\kappa b}\partial_{[\kappa} \hat h_{\lambda]}^e
+2\delta^{\rho[a}\partial_{[\mu} \hat h_{\rho]}^{c]}
\delta_{b e}\hat h^{\lambda c}\delta^{\kappa b}\partial_{[\kappa} \hat h_{\lambda]}^e
+2\delta^{\rho[a}\partial_{[\mu} \hat h_{\rho]}^{c]}
\delta_{b e}\delta^{\lambda c}\hat h^{\kappa b}\partial_{[\kappa} \hat h_{\lambda]}^e
\right.\nonumber\\&&\left.
+\partial^{[c} \hat h^{a]}_\mu\hat h_{b e}\delta^{\lambda c}\delta^{\kappa b}\partial_{[\kappa} \hat h_{\lambda]}^e
+\partial^{[c} \hat h^{a]}_\mu\delta_{b e}\hat h^{\lambda c}\delta^{\kappa b}\partial_{[\kappa} \hat h_{\lambda]}^e
+\partial^{[c} \hat h^{a]}_\mu\delta_{b e}\delta^{\lambda c}\hat h^{\kappa b}\partial_{[\kappa} \hat h_{\lambda]}^e
-\mu \leftrightarrow b\right]
\nonumber\\&&
+\hat h^\nu_b \left[\partial_\mu \left(2 \hat h^{\rho[a}\partial_{[\nu} \hat h_{\rho]}^{b]}
+\hat h_{\nu d}\delta^{\rho a}\delta^{\sigma b}\partial_{[\sigma}\hat h_{\rho]}^d 
+\delta_{\nu d}\hat h^{\rho a}\delta^{\sigma b}\partial_{[\sigma}\hat h_{\rho]}^d 
+\delta_{\nu d}\delta^{\rho a}\hat h^{\sigma b}\partial_{[\sigma}\hat h_{\rho]}^d \right)
+4 \delta^{\rho[a}\partial_{[\mu} \hat h_{\rho]}^{c]}
\delta^{\lambda[c}\partial_{[\nu} \hat h_{\lambda]}^{b]}
\right.\nonumber\\&&\left.
+2 \delta^{\rho[a}\partial_{[\mu} \hat h_{\rho]}^{c]}\partial^{[b} \hat h^{c]}_\nu
+2\partial^{[c} \hat h^{a]}_\mu \delta^{\lambda[c}\partial_{[\nu} \hat h_{\lambda]}^{b]}
+\partial^{[c} \hat h^{a]}_\mu \partial^{[b} \hat h^{c]}_\nu
-\mu \leftrightarrow \nu \right]
\nonumber\\&&
-\frac{1}{2} \left[
\partial_b \left(\hat h_{c e} \hat h^{\sigma b} \delta^{\lambda c}\partial_{[\lambda}\hat h_{\sigma]}^e
+\hat h_{c e} \delta^{\sigma b} \hat h^{\lambda c}\partial_{[\lambda}\hat h_{\sigma]}^e
+\delta_{c e} \hat h^{\sigma b} \hat h^{\lambda c}\partial_{[\lambda}\hat h_{\sigma]}^e \right)
+4 \hat h^{\sigma[b}\partial_{[b} \hat h_{\sigma]}^{d]} \delta^{\kappa[d}\partial_{[c} \hat h_{\kappa]}^{c]}
\right. \nonumber\\ &&\left.
+2 \hat h^{\sigma[b}\partial_{[b} \hat h_{\sigma]}^{d]}\partial^{[c} \hat h^{d]}_c
+2\hat h_{b e}\delta^{\sigma b}\delta^{\lambda d}\partial_{[\lambda} \hat h_{\sigma]}^e
\delta^{\kappa[d}\partial_{[c} \hat h_{\kappa]}^{c]}
+2\delta_{b e}\hat h^{\sigma b}\delta^{\lambda d}\partial_{[\lambda} \hat h_{\sigma]}^e
\delta^{\kappa[d}\partial_{[c} \hat h_{\kappa]}^{c]}
+2\delta_{b e}\delta^{\sigma b}\hat h^{\lambda d}\partial_{[\lambda} \hat h_{\sigma]}^e
\delta^{\kappa[d}\partial_{[c} \hat h_{\kappa]}^{c]}
\right. \nonumber\\ &&\left.
+\hat h_{b e}\delta^{\sigma b}\delta^{\lambda d}\partial_{[\lambda} \hat h_{\sigma]}^e\partial^{[c} \hat h^{d]}_c
+\delta_{b e}\hat h^{\sigma b}\delta^{\lambda d}\partial_{[\lambda} \hat h_{\sigma]}^e\partial^{[c} \hat h^{d]}_c
+\delta_{b e}\delta^{\sigma b}\hat h^{\lambda d}\partial_{[\lambda} \hat h_{\sigma]}^e\partial^{[c} \hat h^{d]}_c
+4\delta^{\sigma[b}\partial_{[b} \hat h_{\sigma]}^{d]}\hat h^{\kappa[d}\partial_{[c} \hat h_{\kappa]}^{c]}
\right. \nonumber\\ &&\left.
+2 \partial^{[d} \hat h^{b]}_b
\hat h^{\kappa[d}\partial_{[c} \hat h_{\kappa]}^{c]}
+2\delta^{\sigma[b}\partial_{[b} \hat h_{\sigma]}^{d]}
\hat h_{c f} \delta^{\kappa d} \delta^{\tau c}\partial_{[\tau} \hat h_{\kappa]}^f
+2\delta^{\sigma[b}\partial_{[b} \hat h_{\sigma]}^{d]}
\delta_{c f} \hat h^{\kappa d} \delta^{\tau c}\partial_{[\tau} \hat h_{\kappa]}^f
+2\delta^{\sigma[b}\partial_{[b} \hat h_{\sigma]}^{d]}
\delta_{c f} \delta^{\kappa d} \hat h^{\tau c}\partial_{[\tau} \hat h_{\kappa]}^f
\right. \nonumber\\ &&\left.
+\partial^{[d} \hat h^{b]}_b \hat h_{c f} \delta^{\kappa d} \delta^{\tau c}\partial_{[\tau} \hat h_{\kappa]}^f
+\partial^{[d} \hat h^{b]}_b \delta_{c f} \hat h^{\kappa d} \delta^{\tau c}\partial_{[\tau} \hat h_{\kappa]}^f
+\partial^{[d} \hat h^{b]}_b \delta_{c f} \delta^{\kappa d} \hat h^{\tau c}\partial_{[\tau} \hat h_{\kappa]}^f
-b \leftrightarrow c \right] \delta_\mu^a 
\nonumber\\&&
-\frac{1}{2} \hat h_b^\nu \left[
\partial_\nu \left(2 \hat h^{\sigma[b}\partial_{[c} \hat h_{\sigma]}^{c]}
+\hat h_{c e} \delta^{\sigma b} \delta^{\lambda c}\partial_{[\lambda}\hat h_{\sigma]}^e
+\delta_{c e} \hat h^{\sigma b} \delta^{\lambda c}\partial_{[\lambda}\hat h_{\sigma]}^e
+\delta_{c e} \delta^{\sigma b} \hat h^{\lambda c}\partial_{[\lambda}\hat h_{\sigma]}^e \right)
+4\delta^{\sigma[b}\partial_{[\nu} \hat h_{\sigma]}^{d]}
\delta^{\kappa[d}\partial_{[c} \hat h_{\kappa]}^{c]}
\right.\nonumber\\&&\left.
+2\delta^{\sigma[b}\partial_{[\nu} \hat h_{\sigma]}^{d]}\partial^{[c} \hat h^{d]}_c
+2\partial^{[d} \hat h^{b]}_\nu\delta^{\kappa[d}\partial_{[c} \hat h_{\kappa]}^{c]}
+\partial^{[d} \hat h^{b]}_\nu \partial^{[c} \hat h^{d]}_c
-\nu \leftrightarrow c \right] \delta_\mu^a
\nonumber\\&&
-\frac{1}{2} \hat h_c^\rho\left[
\partial_b \left(2 \hat h^{\sigma[b}\partial_{[\rho} \hat h_{\sigma]}^{c]}
+\hat h_{\rho e} \delta^{\sigma b} \delta^{\lambda c}\partial_{[\lambda}\hat h_{\sigma]}^e
+\delta_{\rho e} \hat h^{\sigma b} \delta^{\lambda c}\partial_{[\lambda}\hat h_{\sigma]}^e
+\delta_{\rho e} \delta^{\sigma b} \hat h^{\lambda c}\partial_{[\lambda}\hat h_{\sigma]}^e \right)
+4\delta^{\sigma[b}\partial_{[b} \hat h_{\sigma]}^{d]}
\delta^{\kappa[d}\partial_{[\rho} \hat h_{\kappa]}^{c]}
\right.\nonumber\\&&\left.
+2\delta^{\sigma[b}\partial_{[b} \hat h_{\sigma]}^{d]}\partial^{[c} \hat h^{d]}_\rho
+2\partial^{[d} \hat h^{b]}_b
\delta^{\kappa[d}\partial_{[\rho} \hat h_{\kappa]}^{c]}
+\partial^{[d} \hat h^{b]}_b \partial^{[c} \hat h^{d]}_\rho
-b \leftrightarrow \rho \right] \delta_\mu^a
\nonumber\\
&&-\frac{1}{2} \left[
\partial_b \left(2 \hat h^{\sigma[b}\partial_{[c} \hat h_{\sigma]}^{c]}
+\hat h_{c e} \delta^{\sigma b} \delta^{\lambda c}\partial_{[\lambda}\hat h_{\sigma]}^e
+\delta_{c e} \hat h^{\sigma b} \delta^{\lambda c}\partial_{[\lambda}\hat h_{\sigma]}^e
+\delta_{c e} \delta^{\sigma b} \hat h^{\lambda c}\partial_{[\lambda}\hat h_{\sigma]}^e \right)
+4 \delta^{\sigma[b}\partial_{[b} \hat h_{\sigma]}^{d]}
\delta^{\kappa[d}\partial_{[c} \hat h_{\kappa]}^{c]}
\right.\nonumber \\ &&\left.
+2 \delta^{\sigma[b}\partial_{[b} \hat h_{\sigma]}^{d]}\partial^{[c} \hat h^{d]}_c
+2\partial^{[d} \hat h^{b]}_b \delta^{\kappa[d}\partial_{[c} \hat h_{\kappa]}^{c]}
+\partial^{[d} \hat h^{b]}_b \partial^{[c} \hat h^{d]}_c
-b \leftrightarrow c \right] \hat h_\mu^a
\nonumber\\&&
-\frac{1}{2} \hat h_b^\nu \hat h_c^\rho\left[
\partial_\nu \left(2 \delta^{\sigma[b}\partial_{[\rho} \hat h_{\sigma]}^{c]}
+\partial^{[c}\hat h^{b]}_\rho \right)
-\nu \leftrightarrow \rho \right] \delta_\mu^a
-\frac{1}{2} \hat h_c^\rho\left[
\partial_b \left(2 \delta^{\sigma[b}\partial_{[\rho} \hat h_{\sigma]}^{c]}
+\partial^{[c}\hat h^{b]}_\rho \right)
-b \leftrightarrow \rho \right] \hat h_\mu^a
\nonumber\\&&
-\frac{1}{2} \hat h_b^\nu \left[
\partial_\nu \left(2 \delta^{\sigma[b}\partial_{[c} \hat h_{\sigma]}^{c]}
+\partial^{[c}\hat h^{b]}_c \right)
-\nu \leftrightarrow c \right]\hat h_\mu^a.
\label{Gterm_h3}
\end{eqnarray}
Within the expressions ($\ref{Gterm_h1}$), ($\ref{Gterm_h2}$) and ($\ref{Gterm_h3}$) the Kronecker symbols
have only been contracted, where this contraction has improved the representation. 
The expectation values of the terms of the Einstein tensor which are linear in $\hat h^\mu_m$ and quadratic
in $\hat h^\mu_m$ yield no additional terms arising from the noncommutativity of the tetrad field. This means that
it holds

\begin{eqnarray}
\langle \Phi|G_\mu^m[\hat h^1]|\Phi \rangle=G_\mu^m[h^1]\quad,\quad
\langle \Phi|G_\mu^m[\hat h^2]|\Phi \rangle=G_\mu^m[h^2]. 
\label{expectation_value_Gh1_Gh2}
\end{eqnarray}
In case of the term $\hat G_\mu^m[\hat h^1]$ this is obvious, since no permutation of $\hat H_\mu^m$ and
$\hat H_\mu^{m \dagger}$ has to be performed. In case of $\hat G_\mu^m[\hat h^2]$ this property arises from
the fact that there exist only terms which contain derivatives of $\hat h^\mu_m$ or $\hat H_\mu^m$ and
according to ($\ref{commutator_derivatives_H}$) the commutators lead to derivatives acting on the other factors
of the corresponding term. Since these terms contain no further fields on the one hand and they are just quadratic
in $H_\mu^m$ on the other hand, the commutators vanish. From ($\ref{identity_quadratic}$) one can also see that 
there arise no additional terms from the expressions quadratic in $\hat h^\mu_m$, if there appear no further
field factors.
The expectation value of the term of the Einstein tensor depending on $\hat h^\mu_m$ to the third order,
$\hat G_\mu^m[\hat h^3]$, yields additional terms depending on the noncommutativity parameter. To calculate
the expectation value the identities ($\ref{identity_third_power}$) and ($\ref{identity_third_power_second}$)
have to be used. This leads to the following expression:

\begin{eqnarray}
&&\langle \Phi|\hat G_\mu^a[\hat h^3]|\Phi \rangle
=\left[\partial_\mu \left(\delta^{\sigma b} \Xi_{d \ \ b\ [\sigma\rho]}^{\ ad\ \rho}
+\delta^{\rho a}\Xi_{d \ \ b\ [\sigma\rho]}^{\ bd\ \sigma} 
+\delta_{bd}\Xi_{\ \ \ \ \ \ [\sigma\rho]}^{abd\rho\sigma}\right)
+4 \delta^{\lambda[c}\Omega^{[ac]b]\rho}_{\ \ \ \ \ \ \ [\mu\rho][b\lambda]}
+2 \Omega^{[ac]\ \ \rho\ \ \ \ [bc]}_{\ \ \ \ b\ \ [\mu\rho]}
\right.\nonumber\\&&\left.
+2\delta^{\rho a}\delta^{\sigma c}\delta^{\lambda[c}\Omega^{\ db] \ \ \ \ \ }_{d \ \ \ \mu[\sigma\rho][b\lambda]}
+2\delta_{\mu d}\delta^{\sigma c}\delta^{\lambda[c}\Omega^{adb]\rho\ \ \ \ }_{\ \ \ \ \ [\sigma\rho][b\lambda]}
+2\delta_{\mu d}\delta^{\rho a}\delta^{\lambda[c}\Omega^{cdb]\sigma\ \ \ \ }_{\ \ \ \ \ [\sigma\rho][b\lambda]}
+\delta^{\rho a}\delta^{\sigma c}\Omega^{\ d\ \ \ \ \ \ [bc]}_{d\ b\mu[\sigma\rho]}
\right.\nonumber\\&&\left.
+\delta_{\mu d}\delta^{\sigma c}\Omega^{ad\ \rho\ \ \ [bc]}_{\ \ b\ [\sigma\rho]}
+\delta_{\mu d}\delta^{\rho a}\Omega^{cd\ \sigma\ \ \ [bc]}_{\ \ b\ [\sigma\rho]}
+4 \delta^{\rho[a} \tilde \Omega^{c][c b]\ \ \ \lambda}_{\ \ \ \ \ [\mu\rho]\ [b\lambda]}
+2 \tilde \Omega^{\ [cb][ca]\lambda\ \ }_{\mu\ \ \ \ \ \ \ \ [b\lambda]}
+2 \delta^{\rho[a}\delta^{\lambda c}\delta^{\kappa b}\tilde \Omega^{c]\ e}_{\ \ e \ [\mu\rho]b[\kappa\lambda]}
\right.\nonumber\\&&\left.
+2 \delta^{\rho[a}\delta_{b e}\delta^{\kappa b}\tilde \Omega^{c]ce\ \ \ \ \lambda\ \ }_{\ \ \ \ [\mu\rho]\ [\kappa\lambda]}
+2 \delta^{\rho[a}\delta_{b e}\delta^{\lambda c}\tilde \Omega^{c]be\ \ \ \ \kappa\ \ }_{\ \ \ \ [\mu\rho]\ [\kappa\lambda]}
+\delta^{\lambda c} \delta^{\kappa b}\tilde \Omega^{\ \ \ e[ca]}_{\mu e \ \ \ \ \ b[\kappa\lambda]}
+\delta_{be}\delta^{\kappa b}
\tilde \Omega^{\ ce[ca]\lambda\ \ }_{\mu\ \ \ \ \ \ \ [\kappa\lambda]}
+\delta_{be}\delta^{\lambda c}\tilde \Omega^{\ be[ca]\kappa\ \ }_{\mu\ \ \ \ \ \ \ [\kappa\lambda]}
-\mu \leftrightarrow b\right]\nonumber\\&&
+\left[
\Omega^{\ [ab]\nu\ \rho}_{b \ \ \ \ \ \mu\ [\nu\rho]}+\Delta^{\ [ab]\nu\rho}_{b \ \ \ \ \ \ \mu[\nu\rho]}
+\delta^{\rho a}\delta^{\sigma b}\Omega^{\ \ d\nu\ \ }_{bd \ \ \ \mu\nu[\sigma\rho]}
+\delta^{\rho a}\delta^{\sigma b}\Delta^{\ \ d\nu\ }_{bd \ \ \ \nu\mu[\sigma\rho]}
+\delta_{\nu d}\delta^{\sigma b}\Omega^{\ ad\nu\ \rho}_{b \ \ \ \mu\ [\sigma\rho]}
+\delta_{\nu d}\delta^{\sigma b}\Delta^{\ ad\nu\rho}_{b \ \ \ \ \ \mu[\sigma\rho]}
\right.\nonumber\\&&\left.
+\delta_{\nu d}\delta^{\rho a}\Omega^{\ bd\nu\ \sigma}_{b \ \ \ \mu\ [\sigma\rho]}
+\delta_{\nu d}\delta^{\rho a}\Delta^{\ bd\nu\sigma}_{b \ \ \ \ \ \mu[\sigma\rho]}
+4\delta^{\rho[a}\delta^{\lambda[c}
\Omega^{\ c]b]\nu\ \ \ \ }_{b\ \ \ \ [\mu\rho][\nu\lambda]}
+2\delta^{\rho[a}\Omega^{\ c]\ \nu\ \ \ \ [bc]}_{b\ \ \nu\ [\mu\rho]}
+2\delta^{\lambda[c}\Omega^{\ \ b]\nu[ca]}_{b\mu\ \ \ \ \ \ [\nu\lambda]}
+\Omega^{\ \ \ \ \nu[ca][bc]}_{b\mu\nu}
-\mu \leftrightarrow \nu \right]
\nonumber\\&&
-\frac{1}{2}\left[\partial_b \left(
\delta^{\lambda c} \Xi^{\ be\ \sigma}_{e\ \ c\ [\lambda\sigma]}
+\delta^{\sigma b} \Xi^{\ ce\ \lambda}_{e\ \ c\ [\lambda\sigma]}
+\delta_{ce} \Xi^{bce\sigma\lambda}_{\ \ \ \ \ \ [\lambda\sigma]}\right)
+4 \delta^{\kappa[d}\Omega^{[bd]c]\sigma}_{\ \ \ \ \ \ \ [b\sigma][c\kappa]}
+2 \Omega^{[bd]\ \sigma\ \ \ [cd]}_{\ \ \ \ c\ [b\sigma]}
\right. \nonumber\\ &&\left.
+2\delta^{\sigma b}\delta^{\lambda d}\delta^{\kappa[d}\Omega^{\ ec]}_{e\ \ b[\lambda\sigma][c\kappa]}
+2\delta_{b e}\delta^{\lambda d}\delta^{\kappa[d}\Omega^{bec]\sigma}_{\ \ \ \ \ [\lambda\sigma][c\kappa]}
+2\delta_{b e}\delta^{\sigma b}\delta^{\kappa[d}\Omega^{dec]\lambda}_{\ \ \ \ \ [\lambda\sigma][c\kappa]}
+\delta^{\sigma b}\delta^{\lambda d}\Omega^{\ e\ \ \ \ \ \ [cd]}_{e\ cb[\lambda\sigma]}
\right. \nonumber\\ &&\left.
+\delta_{be}\delta^{\lambda d}\Omega^{be\ \sigma\ \ \ \ [cd]}_{\ \ c\ \ [\lambda\sigma]}
+\delta_{be}\delta^{\sigma b}\Omega^{de\ \lambda\ \ \ \ [cd]}_{\ \ c\ \ [\lambda\sigma]}
+4\delta^{\sigma[b}\tilde \Omega^{d][dc]\ \ \ \kappa}_{\ \ \ \ \ [b\sigma]\ [c\kappa]}
+2\tilde \Omega^{\ [dc][db]\kappa}_{b\ \ \ \ \ \ \ \ \ [c\kappa]}
\right. \nonumber\\ &&\left.
+2\delta^{\sigma[b}\delta^{\kappa d}\delta^{\tau c}\tilde \Omega^{d]\ f}_{\ \ f \ [b\sigma]c[\tau\kappa]}
+2\delta^{\sigma[b}\delta_{c f}\delta^{\tau c}\tilde \Omega^{d]df\ \ \ \kappa}_{\ \ \ [b\sigma]\ [\tau\kappa]} 
+2\delta^{\sigma[b}\delta_{c f}\delta^{\kappa d}\tilde \Omega^{d]cf\ \ \ \tau}_{\ \ \ \ [b\sigma]\ [\tau\kappa]} 
\right. \nonumber\\ &&\left.
+\delta^{\kappa d} \delta^{\tau c}\tilde \Omega^{\ \ \ f[db]}_{b f \ \ \ \ \ c[\tau\kappa]}
+\delta_{c f}\delta^{\tau c}\tilde \Omega^{\ df[db]\kappa}_{b\ \ \ \ \ \ \ [\tau\kappa]}
+\delta_{c f}\delta^{\kappa d}\tilde \Omega^{\ cf[db]\tau}_{b\ \ \ \ \ \ \ [\tau\kappa]}
-b \leftrightarrow c \right] \delta_\mu^a
\nonumber\\&&
-\frac{1}{2} \left[
2\Omega^{\ [bc]\nu\ \sigma}_{b\ \ \ \ \nu\ [c\sigma]}
+2\Delta^{\ [bc]\nu\sigma}_{b\ \ \ \ \ \ \nu[c\sigma]}
+\delta^{\sigma b}\delta^{\lambda c}\Omega^{\ \ e\nu\ \ }_{be \ \ \nu c[\lambda\sigma]}
+\delta^{\sigma b}\delta^{\lambda c}\Delta^{\ \ e\nu\ \ }_{be \ \ c\nu[\lambda\sigma]}
+\delta_{ce}\delta^{\lambda c}\Omega^{\ be\nu\ \sigma}_{b\ \ \ \nu\ [\lambda\sigma]}
+\delta_{ce}\delta^{\lambda c}\Delta^{\ be\nu\sigma}_{b\ \ \ \ \ \nu[\lambda\sigma]}
\right.\nonumber\\&&\left.
+\delta_{ce}\delta^{\sigma b}\Omega^{\ ce\nu\ \lambda}_{b\ \ \ \nu\ [\lambda\sigma]}
+\delta_{ce}\delta^{\sigma b}\Delta^{\ ce\nu\lambda}_{b\ \ \ \ \ \nu[\lambda\sigma]}
+4\delta^{\sigma[b}\delta^{\kappa[d}
\Omega^{\ d]c]\nu}_{b\ \ \ \ [\nu\sigma][c\kappa]}
+2\delta^{\sigma[b}\Omega^{\ d]\ \nu\ \ \ \ [cd]}_{b\ \ c\ [\nu\sigma]}
\right.\nonumber\\&&\left.
+2\delta^{\kappa[d}
\Omega^{\ \ c]\nu[db]}_{b\nu\ \ \ \ \ \ [c\kappa]}
+\Omega^{\ \ \ \nu[db][cd]}_{b\nu c}
-\nu \leftrightarrow c \right] \delta_\mu^a
\nonumber\\&&
-\frac{1}{2} \left[
2\Omega^{\ [bc]\rho\ \sigma}_{c\ \ \ \ \ b\ [\rho\sigma]}
+2\Delta^{\ [bc]\rho\sigma}_{c\ \ \ \ \ \ b[\rho\sigma]}
+\delta^{\sigma b}\delta^{\lambda c}\Omega^{\ \ e\rho\ \ }_{ce \ \ b\rho[\lambda\sigma]}
+\delta^{\sigma b}\delta^{\lambda c}\Delta^{\ \ e\rho\ \ }_{ce \ \ \rho b[\lambda\sigma]}
+\delta_{\rho e}\delta^{\lambda c}\Omega^{\ be\rho\ \sigma}_{c\ \ \ b\ [\lambda\sigma]}
+\delta_{\rho e}\delta^{\lambda c}\Delta^{\ be\rho\sigma}_{c\ \ \ \ \ b[\lambda\sigma]}
\right.\nonumber\\&&\left.
+\delta_{\rho e}\delta^{\sigma b}\Omega^{\ ce\rho\ \lambda}_{c\ \ \ b\ [\lambda\sigma]}
+\delta_{\rho e}\delta^{\sigma b}\Delta^{\ ce\rho\lambda}_{c\ \ \ \ \ b[\lambda\sigma]}
+4\delta^{\sigma[b}\delta^{\kappa[d}
\Omega^{\ d]c]\rho}_{c\ \ \ \ \ [b\sigma][\rho\kappa]}
+2\delta^{\sigma[b}\Omega^{\ d]\ \rho\ \ \ \ [cd]}_{c\ \ \rho\ [b\sigma]}
\right.\nonumber\\&&\left.
+2\delta^{\kappa[d}\Omega^{\ \ c]\rho[db]}_{cb\ \ \ \ \ \ [\rho\kappa]}
+\Omega^{\ \ \ \rho[db][cd]}_{cb\rho}
-b \leftrightarrow \rho \right] \delta_\mu^a
\nonumber\\&&
-\frac{1}{2} \left[
2\overline \Omega^{[bc]a\ \sigma}_{\ \ \ \ \ b\ [c\sigma]\mu}
+2\tilde \Delta^{[bc]a\sigma}_{\ \ \ \ \ \ b[c\sigma]\mu}
+\delta^{\sigma b}\delta^{\lambda c}\overline \Omega^{\ ea\ \ }_{e\ \ bc[\lambda\sigma]\mu}
+\delta^{\sigma b}\delta^{\lambda c}\tilde \Delta^{\ ea\ \ }_{e\ \ cb[\lambda\sigma]\mu}
+\delta_{c e}\delta^{\lambda c}\overline \Omega^{bea\ \sigma}_{\ \ \ \ b\ [\lambda\sigma]\mu}
+\delta_{c e}\delta^{\lambda c}\tilde \Delta^{bea\sigma}_{\ \ \ \ \ b[\lambda\sigma]\mu}
\right.\nonumber\\&&\left.
+\delta_{ce}\delta^{\sigma b}\overline \Omega^{cea\ \lambda}_{\ \ \ b\ [\lambda\sigma]\mu}
+\delta_{ce}\delta^{\sigma b}\tilde \Delta^{cea\lambda}_{\ \ \ \ \ b[\lambda\sigma]\mu}
+4\delta^{\sigma[b}\delta^{\kappa[d}
\overline \Omega^{d]c]a}_{\ \ \ \ \ [b\sigma][c\kappa]\mu}
+2\delta^{\sigma[b}\overline \Omega^{d]\ a\ \ \ [cd]}_{\ \ c\ [b\sigma]\ \ \ \mu}
\right.\nonumber\\&&\left.
+\delta^{\kappa[d}\overline \Omega^{\ c]a\nu[db]}_{b\ \ \ \ \ \ \ \ [c\kappa]\mu}
+\overline \Omega^{\ \ a[db][cd]}_{bc\ \ \ \ \ \ \ \ \mu}
-b \leftrightarrow c \right]
\nonumber\\
&&-\frac{1}{2}
\left[
2 \delta^{\sigma[b}\Delta^{\ \ c]\nu\rho}_{bc\ \ \ \ \nu[\rho\sigma]}
+\Delta^{\ \ \ \nu\rho\ [cb]}_{bc\rho\ \ \nu}
-\nu \leftrightarrow \rho \right] \delta_\mu^a
-\frac{1}{2}
\left[
2 \delta^{\sigma[b}\tilde \Delta^{\ c]a\rho}_{c\ \ \ \ b[\rho\sigma]\mu}
+\tilde \Delta^{\ \ a\rho\ [cb]}_{c\rho\ \ b\ \ \ \mu}
-b \leftrightarrow \rho \right]
\nonumber\\
&&-\frac{1}{2}
\left[
2\delta^{\sigma[b}\tilde \Delta^{\ c]a\nu}_{b\ \ \ \ \nu[c\sigma]\mu}
+\tilde \Delta^{\ \ a\nu\ [cb]}_{bc\ \ \nu\ \ \ \mu}-\nu \leftrightarrow c \right],
\label{expectation_value_Gh3}
\end{eqnarray}
where the quantities $\Xi_{\ \ \ \ \mu\nu\sigma\rho}^{abc}$,
$\Delta_{\ \ \ \ \mu\nu\lambda\sigma\rho}^{abc}$,
$\tilde \Delta_{\ \ \ \ \mu\lambda\sigma\nu\rho}^{abc}$,
$\Omega_{\ \ \ \ \mu\lambda\nu\sigma\rho}^{abc}$,
$\tilde \Omega_{\ \ \ \ \lambda\mu\nu\sigma\rho}^{abc}$
and $\overline \Omega_{\ \ \ \ \lambda\mu\sigma\nu\rho}^{abc}$ are defined as follows:

\begin{eqnarray}
\Xi_{\ \ \ \ \mu\nu\sigma\rho}^{abc}&=&h_\mu^a h_\nu^b \partial_\sigma h_\rho^c 
-i\Gamma_{\mu\nu}\delta^{ab} \partial_\sigma h_\rho^c 
-i\Gamma_{\mu\rho}\delta^{ac} \partial_\sigma h_\nu^b 
-i\Gamma_{\nu\rho}\delta^{bc} \partial_\sigma h_\mu^a,\nonumber\\
\Delta_{\ \ \ \ \mu\nu\lambda\sigma\rho}^{abc}&=&h_\mu^a h_\nu^b \partial_\lambda \partial_\sigma h_\rho^c 
+i\Gamma_{\mu\nu}\delta^{ab} \partial_\lambda \partial_\sigma h_\rho^c 
+i\Gamma_{\mu\rho}\delta^{ac} \partial_\lambda \partial_\sigma h_\nu^b 
+i\Gamma_{\nu\rho}\delta^{bc} \partial_\lambda \partial_\sigma h_\mu^a,\nonumber\\
\tilde \Delta_{\ \ \ \ \mu\lambda\sigma\nu\rho}^{abc}&=&h_\mu^a \partial_\lambda \partial_\sigma h_\nu^b h_\rho^c 
+i\Gamma_{\mu\nu}\delta^{ab} \partial_\lambda \partial_\sigma h_\rho^c 
+i\Gamma_{\mu\rho}\delta^{ac} \partial_\lambda \partial_\sigma h_\nu^b 
+i\Gamma_{\nu\rho}\delta^{bc} \partial_\lambda \partial_\sigma h_\mu^a,\nonumber\\
\Omega_{\ \ \ \ \mu\lambda\nu\sigma\rho}^{abc}&=&h_\mu^a \partial_\lambda h_\nu^b \partial_\sigma h_\rho^c 
+i\Gamma_{\mu\nu}\delta^{ab} \partial_\lambda \partial_\sigma h_\rho^c 
+i\Gamma_{\mu\rho}\delta^{ac} \partial_\lambda \partial_\sigma h_\nu^b 
+i\Gamma_{\nu\rho}\delta^{bc} \partial_\lambda \partial_\sigma h_\mu^a,\nonumber\\
\tilde \Omega_{\ \ \ \ \lambda\mu\nu\sigma\rho}^{abc}&=&\partial_\lambda h_\mu^a h_\nu^b \partial_\sigma h_\rho^c 
+i\Gamma_{\mu\nu}\delta^{ab} \partial_\lambda \partial_\sigma h_\rho^c 
+i\Gamma_{\mu\rho}\delta^{ac} \partial_\lambda \partial_\sigma h_\nu^b 
+i\Gamma_{\nu\rho}\delta^{bc} \partial_\lambda \partial_\sigma h_\mu^a,\nonumber\\
\overline \Omega_{\ \ \ \ \lambda\mu\sigma\nu\rho}^{abc}&=&\partial_\lambda h_\mu^a \partial_\sigma h_\nu^b h_\rho^c 
+i\Gamma_{\mu\nu}\delta^{ab} \partial_\lambda \partial_\sigma h_\rho^c 
+i\Gamma_{\mu\rho}\delta^{ac} \partial_\lambda \partial_\sigma h_\nu^b 
+i\Gamma_{\nu\rho}\delta^{bc} \partial_\lambda \partial_\sigma h_\mu^a.
\end{eqnarray}
This means that the generalized Einstein field equation has been calculated which is given by
($\ref{expectation_value_Einstein_matter_thirdorder}$) with ($\ref{expectation_value_Gh1_Gh2}$) and thus
($\ref{Gterm_h1}$) and ($\ref{Gterm_h2}$) transformed to the corresponding expressions depending on the
usual expansion field $h^\mu_m$ and ($\ref{expectation_value_Gh3}$).
Accordingly generalized dynamics for the matter field coupled to gravity and for the gravitational field
itself have been derived by building expectation values of the corresponding matter field equation and
the Einstein field equation containing the tetrad field operator $\hat e^\mu_m$ with respect to
coherent states referring to creation and annihilation operators being built from the components of
the tetrad field operator $\hat e^\mu_m$.
There appear imaginary components within the expectation values of the field equations of the matter
and the gravitational field. This is because the expressions within the field equations in terms of
the tetrad field operator are not hermitian, although the tetrad field operator is hermitian, since
a function of hermitian operators is not always hermitian as well. This means that the field equations
contain two independent conditions on the dynamical evolution of the matter field and the gravitational
field respectively being related to the real and the imaginary component.

\section{Summary and Discussion}

It has been suggested a noncommutativity algebra for the components of the tetrad field describing the gravitational
field which behaves thus as an operator. This algebra corresponds to the algebra of the canonical case of usual noncommutative geometry referring to the components of the space-time coordinate. The relation between the usual quantization
in quantum mechanics where is postulated a nonvanishing commutator between position and momentum and the concept of noncommutative geometry where are postulated nonvanishing commutation relations between the components of the position vector corresponds to the relation between canonical quantization of general relativity and the noncommutativity of the tetrad field which has been presented in this paper. This is because canonical quantization of general relativity postulates nonvanishing commutation relations between the quantity describing the gravitational field and its canonical conjugated momentum and in the approach of this paper are postulated nonvanishing commutation relations between the components of the quantity describing the gravitational field. In this sense the concept of a description of general relativity
with a noncommutative tetrad field as it is treated in this paper could be considered as the consequence of an additional aspect
of a fundamental quantum theory of gravity with respect to a kind of semiclassical description of general relativity.

To describe the consequences of the noncommutativity of the tetrad field for the description of general relativity it has been
necessary to transfer the coherent state approach to noncommutative geometry to the case of noncommuting components of the
tetrad field and thus to the case of noncommuting field components. The necessity to use the coherent state approach has its origin in the fact that in the star product approach there are treated fields depending on noncommutative coordinates and therefore in this case it is possible to use a Weyl quantization which is a kind of Fourier transformation between commuting and noncommuting coordinates. After the Weyl quantization products of fields depending on the noncommutative coordinates are expressed by a sum of products of these fields depending on the usual coordinates and in this way the generalization of a field theory depending on noncommutative coordinates can be expressed by additional products depending on the noncommutativity parameter. But in the scenario considered in this paper the components of a field themselves, namely of the gravitational
field represented as tetrad field, do not commute and therefore a Weyl quantization cannot be performed.
In the coherent state approach there are defined new expressions for the quantities depending on the noncommutative coordinates
as expectation values between coherent states which are defined with respect to operators constructed from the components of the
space-time coordinate. This procedure can be applied to coordinates as well as to fields. Therefore the coherent state approach has been transferred to the tetrad field in this paper and accordingly the generalized quantities of general relativity depending on the noncommutative tetrad field are defined as expectation values between coherent states which are defined with respect to operators constructed from the components of the tetrad field operator.
Based on this concept the expectation value of the resulting operator of the metric field and of the volume element operator
have been calculated as examples. After this the generalized dynamics of a matter field coupled to gravity have been determined, a fermionic field especially, and the generalized dynamics of the gravitational field itself. These dynamics were obtained by replacing the usual tetrad field by the corresponding tetrad field operator within the actions of the fermionic field coupled to gravity and the Einstein-Hilbert action describing the dynamics of the gravitational field, varying these actions by the tetrad field operator and then building the expectation values between the coherent states.
Since the expressions in an exact calculation would become very large, an expansion of the gravitational field around the Minkowski metric or the corresponding Kronecker symbol describing the corresponding tetrad field respectively was used and a calculation to the third order in the expansion field was performed, since the terms to the first and second order do not lead to a modification of the expressions in case of the Einstein field equation.

The deviation terms of the field equations, which of course depend on the tensor $\Lambda^{\mu\nu}$ defining the noncommutativity algebra of the terad fields, are of lower order in $h^\mu_m$ than the terms from which they arise.
Since they are obtained from the commutator of the tetrad field being proportional to $\Lambda^{\mu\nu}$ the power
is two orders lower. In case of the free Einstein equation this means that the deviation terms are of first order
in $h^\mu_m$ and this implies that even in case of a small $\Lambda^{\mu\nu}$ for a very small perturbation of the 
gravitational field the deviation terms yield a bigger contribution than the usual terms to the second and third order
in $h^\mu_m$. Thus in the generalized Einstein equation it seems to be advantageous to consider only the terms to the first order in $h^\mu_m$ concerning the treatment of further investigations of the theory like the solution of the field equations
for the matter field and the gravitational field or the derivation of a propagator. It would be interesting to explore
the possible relation between the presented theory and the canonical quantization of general relativity.   
Noncommutativity relations between coordinates can be derived as a consequence of a generalized uncertainty relation
between position and momentum in quantum mechanics. Analogously it could make sense to postulate a generalized quantinzation
rule for the gravitational field and its canonical conjugated momentum and to derive the noncommutativity of the tetrad field from such a generalized quantum description of the gravitational field.
\\ \noindent
$Acknowledgement$: I would like to thank the Messer Stiftung for financial support and Piero Nicolini for fruitful discussions.

\end{document}